\documentclass[]{aastex631}

\usepackage[T1]{fontenc} 
\usepackage{graphicx}
\usepackage{dcolumn}
\usepackage{bm}
\usepackage{hyperref}
\usepackage{physics}
\usepackage{slashed}
\usepackage{lmodern}
\usepackage{multirow}
\usepackage{soul}
\usepackage{booktabs}

\received{January 11, 2025}
\revised{November 9, 2025}
\accepted{\today}

\shorttitle{Isolated BBH Formation and Merger Rates}
\shortauthors{Smith \& Kaplinghat}

\begin{document}

\title{Isolated Binary Black Hole Formation and Merger Rates from Galaxy Evolution}

\author[0000-0002-1732-8040]{Tyler B. Smith}
\affiliation{Department of Physics and Astronomy \\
University of California, Irvine, CA 92697, USA}

\author[0000-0001-8555-0164]{Manoj Kaplinghat}
\affiliation{Department of Physics and Astronomy \\
University of California, Irvine, CA 92697, USA}

\begin{abstract}

The LIGO-Virgo-KAGRA (LVK) collaboration has detected over 150 confirmed gravitational wave events through O4a. Binary black hole (BBH) systems represent the overwhelming majority of these observations. We construct a model for the population of the BBHs based on the distribution of metallicities in galaxies and state-of-the-art stellar evolution models implemented through the Stellar EVolution N-body (\texttt{SEVN}) code. We calculate the redshift evolution of the total merger rate of BBHs and the differential rates with respect to primary mass, secondary mass, and the mass ratio. We explore variations in the delay-time distribution's (DTD) power-law index and show that it affects the total merger rate's spectral shape, but primarily acts as an amplitude shift on the differential rates. When comparing to the primary mass distribution, our results indicate that either the average IMF in dwarf galaxies must be top heavy, or most of the 30--40 $\rm M_\odot$ black holes must be formed through a dynamical capture mechanism. For masses greater than about $50 \, \rm M_\odot$, the predicted number of BBH systems plummet to zero, revealing the well-known mass gap due to the pair instability mechanism and mass loss in binary systems. 

\end{abstract}


\section{Introduction} \label{sec:intro}
In 2015, the LIGO collaboration discovered the first direct evidence for gravitational waves (GWs) with the detection of event GW150914 \citep{Abbott2016}. Currently observing run four (O4) is nearing completion, with over 150 confirmed GW events detected by the LIGO--Virgo--KAGRA (LVK) collaboration \citep{GWTC1,Abbott2020c,GWTC3,GWTC4pop} up to O4a. Among the GW events are binary neutron star (BNS), neutron star--black hole (NS--BH), and binary black hole (BBH) mergers encompassing a wide mass spectrum. 

These GW observations have revolutionized our understanding of compact objects. Prior to the first GW detections, observations of black holes in X-ray binaries were largely limited to masses of $\lesssim 20 \, \rm M_\odot$ \citep{Farr:2010tu,Ozel:2010su} which was consistent with the maximum black hole mass calculated in \citep{Fryer2001b}. While observational evidence pointed to an upper mass $\lesssim 30 \, \rm M_\odot$, theoretical modeling of low-metallicity systems asserted a higher threshold around $\sim 80 \, \rm M_\odot$ \citep{belczynski2010ads,Mapelli2009MNRAS}. Thus, the detection of GW150914 \citep{Abbott2016} which featured component masses of $36 \, \rm M_\odot$ and $29 \, \rm M_\odot$, was not a complete surprise \citep{Mandel:2018hfr}, but still challenged existing assumptions primarily related to the lack of observation of such massive black holes and prompted a revision of black hole formation theories. Subsequent merger events have pushed beyond observational gaps, placing increasing pressure on theoretical models.

Two notable examples of changing perspectives are related to the so-called mass gaps in the compact object mass spectrum. The first mass gap, the \textit{lower-mass gap}, occurs between the highest-mass neutron stars and the lowest-mass black holes. Neutron stars have long had a theoretical upper limit of $\sim 2.2 \, \rm M_\odot$, termed the Tolman--Oppenheimer--Volkoff (TOV) limit \citep{Bombaci96,Kalogera:1996ci,Rezzolla:2017aly}. Likewise, black holes were previously thought to only form down to a lower limit of $\sim 5 \, \rm M_\odot$ \citep{Farr:2010tu}, possibly resulting from core-collapse supernova physics \citep{GWTC3pop}. This gap was postulated due to the dearth of observations of black holes below this threshold. Gravitational wave events GW 190814 \citep{LIGOScientific:2020zkf}, GW 200115 \citep{GWTC3,LIGOScientific:2021qlt}, and a first result of the O4 run \citep{LIGOScientific:2024elc}, directly challenge the lower-mass gap expectation. While traditional astrophysical processes can explain the presence of black holes in this range, studies have also shown that lower-mass gap black holes could originate from primordial black holes \citep{Afroz:2024fzp}.

The second mass gap is predicted by the pair instability mechanism \citep{Fowler1964,Barkat1967,Fryer2001,Belczynski2016,Woosley2007,Woosley2017,Woosley2019}, termed the \textit{pair-instability mass gap } or the \textit{upper-mass gap}. When a star's core temperature reaches $\sim 10^9 \, \rm K$ the high-energy photons gain sufficient kinetic energy to annihilate into electron-positron pairs. This conversion reduces the radiation pressure of the star inducing gravitational collapse, leading to ignition of explosive oxygen/silicon burning. For helium core masses $30 \lesssim M_{He}/\rm M_\odot \lesssim 60 $ the core contracts and ignites burning, but it is not sufficient to completely disrupt the star. The star expands and cools to equilibrium, and this process can repeat leading to a series of pulsations lasting from a few hours to $\sim 10,000$ years \citep{Woosley2017}. For higher mass cores $60 \lesssim M_{He}/\rm M_\odot \lesssim 130 $ the explosion disrupts the entire star, resulting in a pair-instability supernova with no remnant. For core masses exceeding this limit, $M_{He} \gtrsim 130 \, \rm M_\odot$, the star collapses directly to a black hole as a result of the pair instability \citep{Heger:2002by}.

That these mass gaps are populated according to LVK observations \citep{Abbott2020a,Abbott2020b,GWTC3,GWTC3pop,GWTC4pop} highlight the increasing discrepancy between current theoretical expectations and observations. There are currently $> 150 $ total confirmed events in O4a, with more to come by the end of the fourth observing run. The new data will shed light on compact objects and may also present new challenges for understanding their nature and origin. In light of this, it is imperative to develop sophisticated models that detail the formation and evolutionary pathways of these compact objects, particularly in the context of mergers. 

In this study, we specifically focus on binary black hole systems that form and evolve through the isolated channel. The isolated pathway consists of binary systems where the component stars interact with each other before merger but have negligible interaction with external pairs or the environment 
\citep{Tutukov1973,Bethe:1998bn,OShaughnessy:2009szr,Dominik:2013tma,Dominik:2014yma,Belczynski2016,Stevenson:2017tfq,Elbert2018,Baibhav:2019gxm,Spera2019,Santoliquido:2020bry,Bouffanais:2021wcr,Broekgaarden:2021hlu, Garcia:2021niy, Stevenson:2022djs, Broekgaarden:2021efa, Iorio2022, Boesky:2024msm}. In contrast, the dynamical channel involves interactions among single or multi-star systems within densely populated stellar environments, such as globular clusters, nuclear star clusters, or young star clusters \citep{Portegies2000,OLeary:2005vqo,Banjeree2010,Farr2017,Mapelli2020b,Santoliquido:2020bry,Gerosa2021,Sedda:2023big}. In the dynamical channel, stars frequently engage in close encounters or exchanges, leading to a series of complex dynamical interactions.

It is common in the literature to adopt a framework based on the use of a population synthesis code coupled with galactic and cosmological relations such as the star formation rate (SFR), galaxy stellar mass function (GSMF), and mass-metallicity relation (MZR) \citep{OShaughnessy:2009szr,Neijssel2019,Santoliquido:2020bry,Bouffanais:2021wcr,Broekgaarden:2021efa,Broekgaarden:2021hlu,Stevenson:2022djs,Boesky:2024msm}. We take a similar approach, utilizing the Stellar EVolution N-body code \texttt{SEVN} \citep{Spera2015,Spera2017,Spera2019,Mapelli2020,Iorio2022}, which incorporates the \textsc{PARSEC} tracks \citep{Bressan2012,Tang2014,Chen2015,Marigo2017,Nguyen2022} as our choice of population synthesis model. We differ from previous studies by adopting an up-to-date, redshift-evolving GSMF based on observational constraints from the COSMOS survey \citep{COSMOS2020}, which provides star-forming measurements out to $z \sim 5.5$. Beyond this range, we extrapolate the observed trend. We note that \citet{Neijssel2019} also consider an evolving GSMF, though theirs is derived from the EAGLE simulations \citep{Furlong2015}. This choice showcases how the number density of binary black holes that merge within the age of the universe evolves over redshift. 

In conjunction with the GSMF, we account for the redshift evolution of the mass-metallicity relation (MZR) by adopting the results of cosmological zoom-in simulations from \citet{Ma:2015ota}. Their results remain consistent with observed MZR trends up to  $z=3$ from \citet{Tremonti2004,Gallazzi2005,Erb2006,Mannucci2009,Zahid2011,Kirby2013,Steidel2014,Sanders2015}. This redshift range covers the bulk of current LVK detections, given the short stellar lifetimes and delay-time distributions of massive binary systems as discussed in detail below. Furthermore, we adopt the star-forming main sequence in \citet{Speagle:2014loa} and constrain the normalization by the core-collapse supernova rate (CCSN) from the ZTF survey \citep{Perley:2020ajb}. 

Our goal is to show how our particular choices of galactic and cosmological parameterizations, grounded in empirical observations, can reproduce the total merger rate and component mass distributions inferred from the LVK collaboration. We highlight how the isolated channel alone cannot reconcile the LVK results, and what it would take from star formation and stellar physics to do so. These results further strengthen the argument for inclusion of the dynamical population, in particular at higher masses. 

While our general approach is common, we further highlight how our study goes beyond previous works. Most of the works on isolated binary evolution cited above have focused primarily on the total merger rate density and its evolution with redshift. Though there have also been efforts to examine specific aspects of the black hole mass distributions. For example, \citet{Broekgaarden:2021efa} investigated the chirp mass distribution inferred from gravitational wave observations, while \citet{Neijssel2019} analyzed the total mass distribution, though without incorporating the effects of pair-instability supernovae. Their justification was that similar high-mass cutoffs can emerge from particular combinations of galaxy and stellar evolution relations. Additionally, \citet{Tanikawa:2021qqi} performed model comparisons against the cosmic merger rate and its mass derivatives across Pop I-III systems. However, unlike our work, \citet{Tanikawa:2021qqi} does not incorporate empirically constrained galaxy-based ingredients such as the evolving GSMF and MZR.

We extend this body of work by modeling not only the total merger rate, but also the differential merger rate density with respect to primary mass, secondary mass, and mass ratio. This enables a more detailed and physically grounded comparison with both current LVK data and future observations. Notably, the secondary mass spectrum remains relatively underexplored, though see recent work from \citet{Sadiq:2023zee,Farah:2023swu}. We also briefly examine the impact of the DTD on the redshift evolution of the merger rate, finding that even modest variations can lead to noticeable differences. While this manuscript was under review, LVK released O4a, providing an updated and expanded catalog of GW events. We include comparisons with these most recent results in this work.

Our paper is structured as follows. In Section \ref{sec:sims}, we begin by discussing the simulations used to inform and constrain our analysis. Following this, we construct the population of black holes in Section \ref{sec:dNdm1}. The calculation of the merger rate density is detailed in Section \ref{sec:mrdcalc}. In Section \ref{sec:results}, we compare the results of our model to those of the LVK collaboration. We conclude in Section \ref{sec:conclusions}.

\section{Methodology}
\subsection{Zero-Age Main Sequence to Compact Object Evolution with \texttt{SEVN} Simulations}
\label{sec:sims}

The first ingredient to properly model the binary black hole population is stellar evolution and dynamics. To encapsulate the complex and broad physical processes involved, we use the stellar evolution code \texttt{SEVN v2.7.3} \citep{Spera2015, Spera2017, Spera2019, Mapelli2020, Iorio2022} to evolve stars from their zero-age main sequence (ZAMS) masses to their final remnant masses, with a focus on binary mergers. We adopt the default parameters unless noted otherwise, e.g. PARSEC tracks with overshooting parameter $\lambda=0.5$, and refer the reader to \citet{Spera2019} and \citet{Iorio2022} for further details on these prescriptions and the effects of modifying them.

\begin{figure}[t!]
    \centering
    \includegraphics[width=\textwidth]{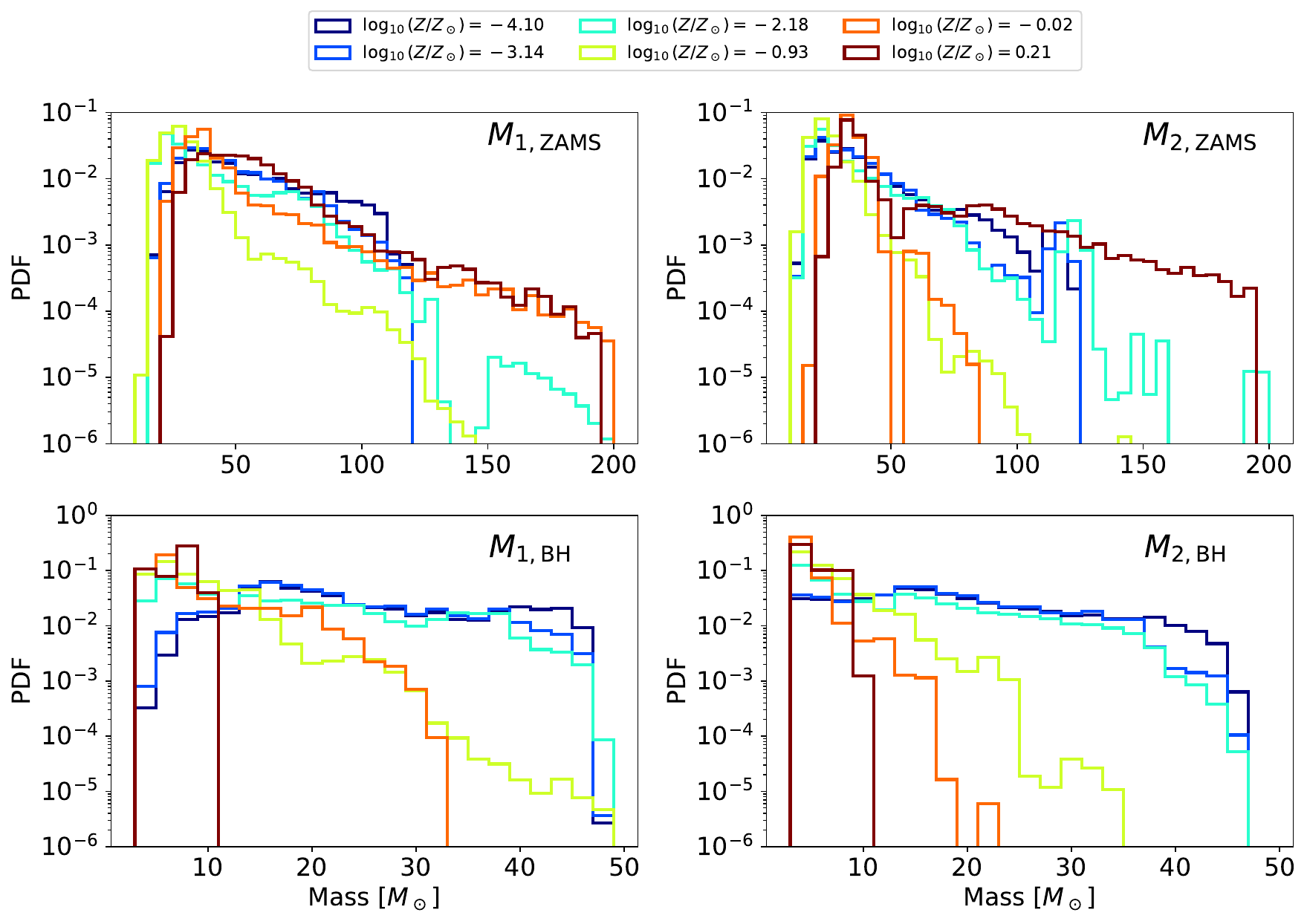}
    \caption{Probability density functions of stellar and remnant masses across metallicity. The top row shows the zero-age main sequence (ZAMS) masses of the primary (left; $M_{1,\mathrm{ZAMS}}$) and secondary (right; $M_{2,\mathrm{ZAMS}}$) stars in merging binary systems, while the bottom row shows the corresponding black hole masses, where $M_{1,\mathrm{BH}}$ denotes the more massive black hole in the binary. Each curve represents a different metallicity, as indicated by the color scheme, and all distributions are weighted by the initial mass function and normalized to highlight relative contributions. }
    \label{fig:histm1m2}
\end{figure}

We evolve $\sim 8 \times 10^7$ binary configurations per 20 distinct metallicities spanning $-4 \leq \log_{10}(Z/Z_\odot) \leq 0.2$. Progenitor masses for the primary star are uniformly sampled in the range $10 \leq M /\rm M_\odot \leq 200$ with a spacing of $1 \rm \, M_\odot$.  To obtain a representative sample of binary systems, following the sampling of $M_{1, \rm ZAMS}$ and $\log_{10}Z$, we sample the mass ratio, eccentricity, and period from distributions following the results in \citep{Sana2012}:

\begin{subequations}
\begin{align}
    P(q_{ZAMS}) &\propto q_{ZAMS}^{-0.1} , & q_{ZAMS} &= \frac{M_2}{M_1}, & q_{ZAMS} &\in [0,1], \label{eq:1a} \\
    P(\mathcal{P}) &\propto \mathcal{P}^{-0.55} , & \mathcal{P} &= \log_{10}(\text{P/day}), & \mathcal{P} &\in [0.15,5.5], \label{eq:1b} \\
    P(e) &\propto e^{-0.42} , &&& e &\in [0,0.9]. \label{eq:1c}
\end{align}
\label{eq:dists}
\end{subequations}

In our model, we assume zero initial spins for the progenitor stars. This assumption is well justified, as the majority of the star's primordial angular momentum is lost during its evolution due to Roche-lobe overflow mass transfer and wind mass loss \citep{Qin:2018vaa,Bavera:2020inc}. These studies show that as the star expands, its angular momentum is transferred to the outer layers, which are subsequently removed through the aforementioned processes. Specifically, \citet{Bavera:2020inc} highlight that the initial angular momentum is mostly lost by the time a star becomes a helium star, and \citet{Qin:2018vaa} emphasize that most of the angular momentum a star might have had is lost during this expansion phase, making the initial spin largely irrelevant for the final spin of the black hole. By assuming zero initial spin we have simplified the model without sacrificing accuracy, as the black hole's spin is dominated by processes such as mass transfer, tidal synchronization, and accretion in the binary system's later evolutionary stages. 

We restrict our analysis to binary systems containing two black holes that merge within a Hubble time ($\sim$ 14 Gyr). Figure~\ref{fig:histm1m2} shows the distributions of both initial stellar masses (ZAMS) and final black hole (BH) masses for our population of binaries. Because we uniformly sample the primary stellar mass across the full range, the systems are weighted by the IMF to produce a physically meaningful distribution. We define the primary black hole mass, $M_{1,\mathrm{BH}}$, as the more massive BH in the binary. For consistency, we define its progenitor, $M_{1,\mathrm{ZAMS}}$, as the primary star that formed this black hole. Each panel in Figure~\ref{fig:histm1m2} includes contributions from six different metallicity bins spanning our full range. The distributions are normalized per metallicity to highlight the contributions across the mass spectrum. 

In the ZAMS distributions (top panels), we observe a decline in the contribution from very massive stars ($\gtrsim 100 \, \rm M_\odot$) as metallicity decreases. This is because low-metallicity stars retain more mass and are more likely to enter the pair-instability regime, preventing black hole formation. In contrast, high-metallicity stars experience strong mass loss through stellar winds, which prevents them from reaching the pair-instability threshold, allowing them to contribute even at the highest initial masses. However, these high-metallicity stars only produce black holes with masses below $\sim 30 \, \rm M_\odot$ due to extensive wind mass loss \citep{Romagnolo:2023dxc}. Low-metallicity stars, on the other hand, contribute more uniformly across the BH mass range and can form BHs up to $\sim 45 \, \rm M_\odot$ in our simulations. 

\begin{figure}[h!]
    \centering
    \includegraphics[width=0.9\textwidth]{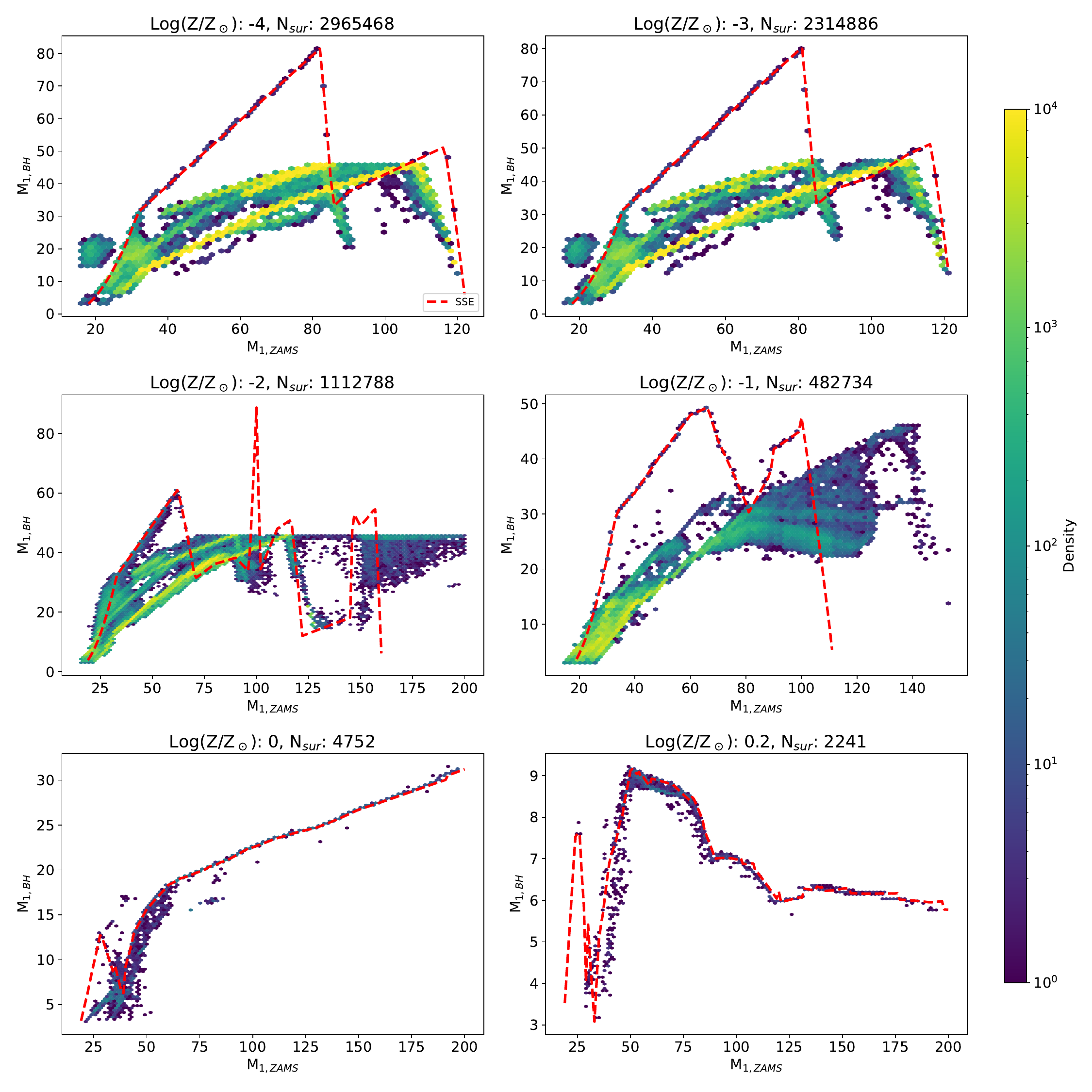}
    \caption{Density plots of primary black hole mass ($M_{1,\mathrm{BH}}$) versus the primary ZAMS mass ($M_{1,\mathrm{ZAMS}}$) for six metallicities. Each panel lists the metallicity and the number of surviving systems ($N_{\mathrm{surv}}$), illustrating the steep decline in BBH formation efficiency toward higher metallicities. The color scale shows the logarithmic density of systems in hexagonal bins of $2\,M_\odot$, while the red dashed lines indicate single-star evolution (SSE) predictions. The initial-final mass relation deviates substantially from the SSE case, reflecting the effects of binarity.} 
    \label{fig:multi_Z_comp}
\end{figure}

To further illustrate the initial-to-final mass relationship, we compare the initial stellar mass to the final black hole mass in Figure~\ref{fig:multi_Z_comp}. This figure is inspired by Figure 5 in \citet{Spera2019}, but differs in key aspects. They examine systems which end up in compact object binaries and find that the majority of binary systems follow the results from single stellar evolution (SSE), while we include the cut $t_{\rm merge} < t_{\rm Hubble}$ and find most of these systems do not track SSE. The red line indicates the SSE case in which each progenitor and metallicity combination produces a unique black hole mass in \texttt{SEVN}. In contrast, the binary stellar evolution (BSE) model exhibits a wide range of black hole masses for each progenitor mass-metallicity combination. Each subplot represents the BSE case via hexagonal binning, width of $\sim 2 \, \rm M_\odot$, where the colorbar logarithmically represents the density. It is well known that merging BBHs have a greater efficiency at low metallicities as is apparent in our figure. Additionally, at intermediate metallicities ($-2 \lesssim \log_{10}(Z/Z_\odot) \lesssim -1$), mass transfer in binaries can produce BHs from progenitors in the mass range that would not do so under SSE. For low-metallicity systems there exists an island above the SSE line at $ M_{\rm ZAMS} \sim 20 \rm M_\odot$. These systems are  predominantly populated by systems where mass ratio reversals takes place, i.e. the secondary progenitor becomes the primary black hole. Mass ratio reversals occur in $\sim 20 \%$ of our merged systems. For the most metal-rich systems, the evolutionary path follows the SSE case, which is due to intense stellar winds disrupting binary systems and leaving only those that evolve in relative isolation.

We note that prescription for the pair-instability mechanism we use is termed M20 in \texttt{SEVN} following the fit from 
\citet{Spera2017} to 1D hydrodynamical simulations from \citet{Woosley2017}. Under this prescription a star will undergo pulsational pair instability if the pre-supernova helium core mass falls between $32 \leq M_{\rm He}/ \rm M_\odot \leq 64$ and a pair-instability supernova for $64 \leq M_{\rm He}/ \rm M_\odot \leq 135$ \citep{Iorio2022}. In the simulations from \citet{Woosley2017} black holes were not found to form between $\sim 50-130 \, \rm M_\odot$, however as is clear from Figure~\ref{fig:multi_Z_comp} we obtain a small number of black holes within this regime. These massive black holes form in the \texttt{SEVN} code as the remnants of low-metallicity progenitor stars of $\sim 100 \rm \, M_\odot$, and avoid the pair instability mechanism during their late dredge-up phase as the mass of the helium core is reduced to just below the threshold required for PISN, $\sim 32 \, \rm M_\odot$ \citep{Iorio2022}. The star can then end its life with roughly the mass at the end of helium depletion phase, producing a massive BH that lies in the pair-instability mass gap. These black holes evolve in relative isolation, mimicking the SSE track, and are exceptionally rare. Thus, predictions for black holes with masses above $\sim 50 \, \rm M_\odot$ should be interpreted with care.

\begin{figure}[t!]
   \centering
    \includegraphics[width=0.5\columnwidth]{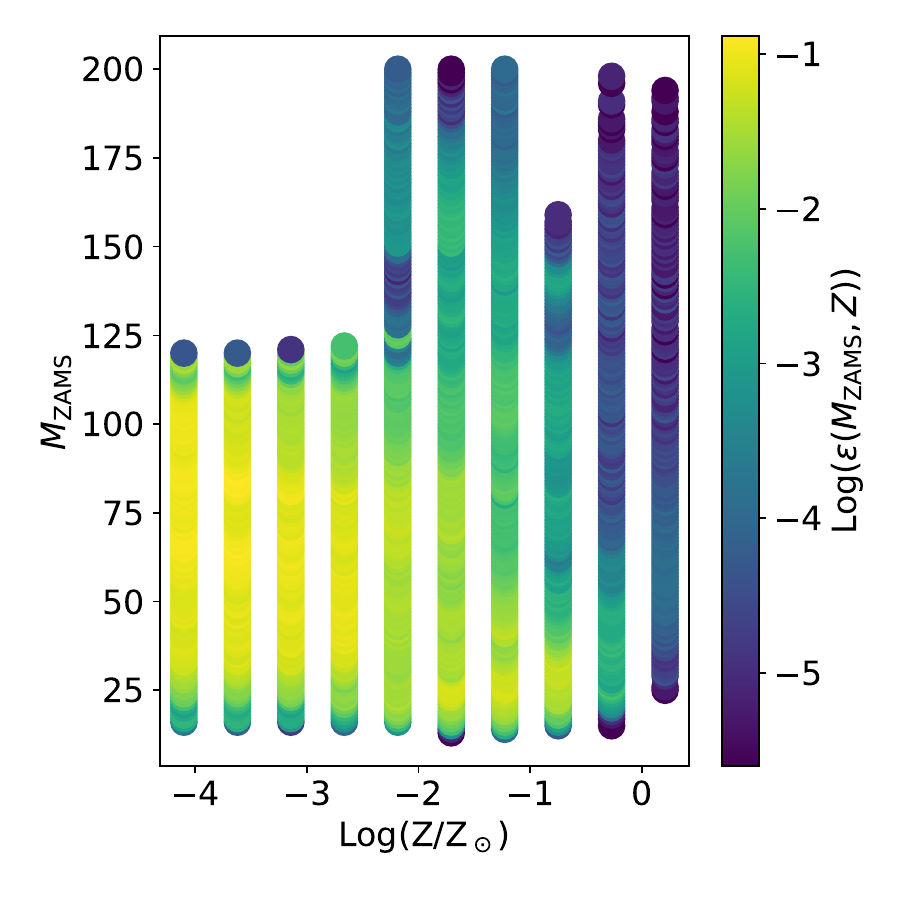}
    \caption{The efficiency for black holes to form given a particular progenitor mass, $M_{\rm ZAMS}$, and metallicity, Z. Low-metallicity progenitors experience a higher efficiency and overall survivability probability as expected. In contrast, higher-metallicity systems, although they may avoid pair instability, have a much lower probability of survival as indicated by the reduced efficiency factor.}
    \label{fig:eff_v_Z}
\end{figure}

It is clear that the mass distribution of surviving binary black hole masses is intricately shaped by the progenitor characteristics and the orbital configurations of each system, in agreement with \citet{Spera2019,Broekgaarden:2021efa,Iorio2022}. To accurately capture this complex relationship, we introduce an efficiency parameter, $\varepsilon(M_{\rm ZAMS},Z)$. This parameter represents the likelihood of a star, for a given progenitor mass and metallicity, to form a black hole of any mass, marginalized over the orbital parameters. This efficiency factor is defined as:

\begin{equation}
    \varepsilon(M_{\rm ZAMS},Z) = \frac{\mathcal{N_{\text{sur}}}(M_{\rm ZAMS},Z)}{\mathcal{N_{\text{sim}}}(M_{\rm ZAMS},Z)}\,,
    \label{eq:eff}
\end{equation}

where $\mathcal{N_{\text{sur}}}(M_{\rm ZAMS},Z)$ represents the number of primary binary black hole progenitors that successfully merge within a Hubble time as determined from our sampling, and $\mathcal{N_{\text{sim}}}(M_{\rm ZAMS},Z)$ denotes the total number of simulated progenitor configurations. The results from Equation \ref{eq:eff} are shown in Figure \ref{fig:eff_v_Z}. This figure demonstrates that stars with lower metallicity have a higher likelihood to form black hole remnants, consistent with findings in various works including, but not limited to \citet{Belczynski2010,mapelli2010,Stevenson:2017tfq, Chruslinska:2018hrb, Giacobbo:2018etu, Neijssel2019, Spera2019, Broekgaarden:2021efa, Schiebelbein-Zwack:2024roj}. While the efficiency factor for creating black holes in general is higher at lower metallicities, the progenitors are also more likely to undergo pair-instability supernova as indicated by the dip in the maximum $M_{\rm ZAMS}$ for black holes at $\log_{10}(Z/Z_\odot) \lesssim -2.5$. This limits the growth and quantity of otherwise fairly massive black holes.

We construct the joint distribution of the primary black hole mass, $M_1$, and mass ratio, $p(M_1, q \mid M_{\rm ZAMS}, Z)$, derived from the population of surviving binary black hole systems. The black hole mass ratio is defined as $q\equiv M_2/M_1$. For each combination of progenitor mass $M_{\rm ZAMS}$ and metallicity $Z$, we use a two-dimensional kernel density estimate over the $M_1$--$q$ space to determine the density. The resulting joint distribution $p(M_1, q \mid M_{\rm ZAMS}, Z)$ is then normalized by our efficiency function from Equation \ref{eq:eff}. Finally, we interpolate the normalized distribution bilinearly over a grid of $Z$ and $M_{\rm ZAMS}$ values, ensuring full representation across the parameter space.

\subsection{Galactic Binary Black Hole Population}
\label{sec:dNdm1}

Building on the results from Section~\ref{sec:sims}, we develop a model for the population of binary black holes that merge within a Hubble time. Our goal is to quantify the growth and abundance of these systems in galaxies of varying mass across cosmic time. The formation rate of such objects is captured by the following integral:

\begin{equation}
    \frac{d^2N}{dM_1 \, dt} (M_\star ,z) = \frac{S(z,M_\star)}{\langle M_{p}\rangle(M_\star,z)} \, \int^{1}_{0} \, dq\int^{Z_{u}}_{Z_{l}} dZ \, \rho (Z | M_\star,z) \int_{M_{ZAMS,\rm low}}^{M_{ZAMS,\rm high}} dM_{\rm ZAMS} \, p(M_{\rm ZAMS}| z, M_\star) \, p(M_1, q|Z,M_{\rm ZAMS}) \,,
    \label{eq:dNdM1dt}
\end{equation}

The limits $Z_{u}$ and $Z_{l}$ are taken from the bounds of our simulations, $Z_{u}=1.5\times 10^{-6}$ and  $Z_{u}=3.2\times 10^{-2}$. The limits of the mass integral are from $M_{\rm ZAMS,\rm low}=8 \, \rm M_\odot$, the mass needed to produce a CCSN, up to $M_{\rm ZAMS,\rm high}=200\, \rm M_\odot$, rather than the Eddington limit, in light of the various stars observed with estimated masses greater than $150 \, \rm M_\odot$ \citep{Crowther2010,bestenlehner2014,bestenlehner2020}. We note that, due to the suppression by the IMF, this change in limits has a negligible impact on the overall result. Although the lower bound does not coincide with the $M_{\rm ZAMS}=10 \, \rm M_\odot$ limit of our simulations, no merging BBHs form with primary masses between $8$–-$10 \, M_\odot$, so the distribution $p(M_1,q\mid M_{\rm ZAMS},Z)$ naturally vanishes in this range. The distribution function of primary black holes, $p(M_1, q \mid M_{\rm ZAMS}, Z)$, is calculated as given in Section \ref{sec:sims}, normalized by $\varepsilon(M_{\rm ZAMS},Z)$ defined in Equation \ref{eq:eff}.

The initial mass function (IMF), i.e., the distribution of stars per unit $M_{\rm ZAMS}$ at a given redshift and galaxy mass is given by $p(M_{\rm ZAMS} | z,M_\star)$. It is defined such when integrated over $M_{\rm ZAMS}$, we get back unity. Here we assume a power-law of the form $M^\xi_{\rm ZAMS}$ for the IMF {\em over the range of stellar masses that gives rise to black holes} and we will constrain this using the core collapse supernova rate. To account for the possibility of a top-heavy IMF, we allow the exponent $\xi$ to change as a function of stellar mass using the relation $M_{p}^{-\alpha + \beta \log_{10}(M_{\rm ZAMS}/8)}$. Here, $\beta$ sets the stellar mass-dependent flattening or steepening of the IMF at higher masses. We choose $\beta$ such that it enhances the high-mass tail while remaining consistent with observational constraints, such as those found in \citet{Marks:2012ia} for globular clusters.

In this work, we present results based on two IMFs, the constant slope of -2.3 (i.e., $\beta =0$) consistent with the Kroupa IMF \citep{Kroupa2001} and the running-slope model with $\beta =0.3$. Note that we choose a slope of $-2.3$ at $8 \rm M_\odot$
for both models to facilitate comparison between the two models. We have not performed a scan of parameters to determine what would best fit the LIGO data, as this exercise would be premature given the uncertainties that still need to be explored. This comparison also highlights the potential impact of metallicity-driven variations of the IMF, the black hole mass distribution, and the number density. There is evidence suggesting potential variations at the high-mass end due to environmental factors such as metallicity in dwarf galaxies and globular clusters \citep{Marks:2012ia,Geha2013,Gennaro18,Weatherford:2021zdf}, as well as work that has explored variations in the IMF and highlighted its impact \citep{Chruslinka2020,Tanikawa:2021qqi}. In particular, the work by \citet{Tanikawa:2021qqi} also examine a top-heavy IMF, $\propto M_{\rm ZAMS}^{-1}$, however only for very metal poor stars, those outside the metallicities we consider here. In contrast, we gradually flatten ours across the mass space for stars large enough to end their lives in a core-collapse supernova.

The metallicity distribution function, $\rho(Z \mid M_\star, z)$, is assumed to follow a log-normal distribution, though this is shown to be a simplifying assumption \citep{vanSon:2022ylf}, within the limits defined by the range of our simulations $-4.1 < \log_{10}(Z/Z_\odot) < 0.2$, and evolves over cosmic time \citep{Zahid2013}. The mean, $\langle Z(M_\star,z) \rangle$, is obtained from \citet{Ma:2015ota}, who fit to high-resolution cosmological zoom-in simulations of the evolution of galaxy mass-metallicity relations, in agreement with results up to z=3 including  \citet{Tremonti2004,Gallazzi2005,Erb2006,Mannucci2009,Zahid2011,Kirby2013,Steidel2014,Sanders2015}. We use the gas-phase metallicity, with the functional form restated here for convenience:

\begin{equation}
    \log\left(\frac{Z_{\rm gas}}{Z_\odot}\right) = 0.35\left[\log\left(\frac{M_\star}{M_\odot}\right)-10\right] + 0.93\,e^{-0.43z} - 1.05 \,. 
\end{equation}

For the variance used in the $\rho(Z \mid M_\star,z)$, we use $\sigma(M_\star,z)=0.4$, constant across redshift, which is the median value found in the compilation of low redshift stellar metallicity observations found in \citet{Simon2019} and the values found in \citet{Gallazzi2005} which have a mean redshift value of $z=0.13$. If instead we chose to use a value akin to that used in \citet{Santoliquido:2020bry}, i.e. $\sigma(M_\star,z)=0.2$, the total binary black hole merger rate varies by less than a factor of 2 overall.

The function outside the integrand in Equation~\ref{eq:dNdM1dt}, $S(z,M_\star)$, represents the total stellar mass formed per year in a galaxy of stellar mass $M_\star$ at redshift $z$, often referred to as the \emph{star-forming main sequence}. We adopt the empirical relation from \citet{Speagle:2014loa}, constructed from a meta-analysis of 25 previous studies. In their formulation, the main sequence is expressed as a function of the cosmic time $t$ (in~Gyr), which we convert to redshift using $t = t(z)$. This choice, when convolved with our GSMF, yields a star-formation rate density which we compare to observational works in Appendix~\ref{app:mssfr}. For convenience, the functional form is given here:

\begin{equation}
\mathrm{SFR}(M_\star, t(z))
= 10^{-(6.51 - 0.11\,t(z))} 
\left(\frac{M_\star}{M_\odot}\right)^{\,0.84 - 0.026\,t(z)}
\; M_\odot\,\mathrm{yr^{-1}}.
\end{equation}

In Equation~\ref{eq:dNdM1dt}, we note that the average progenitor mass is defined with respect to the IMF defined over the entire star-forming range, i.e., $\langle M_{\rm ZAMS}\rangle(M_\star,z) = \int dM_{\rm ZAMS} \, M_{\rm ZAMS} \, p(M_{\rm ZAMS}|M_\star,z)$.

If we write $p(M_{\rm ZAMS}|z, M_\star) = C M^\xi_{\rm ZAMS}$ then we find that 

\begin{equation}
    \frac{C}{\langle M_{\rm ZAMS}\rangle(M_\star,z)} 
    = \frac{f_{bin} K^{CC}}{2 \int^{m_u^{CC}}_{m_l^{CC}} dM_{\rm ZAMS} \, M^\xi_{\rm ZAMS} }\,,
    \label{eq:KCC}
\end{equation}

where $K^{CC}$ is the number of supernova formed per unit stellar mass as given in \citep{Botticella2008} and we assume that this is constant. The limits in the integral are from $m_u^{CC}=8 \, \rm M_\odot$ to $m_u^{CC}=200 \, \rm M_\odot$. The volumetric supernova rate is then given by $R^{CC}(z) = K^{CC} \psi(z)$. We use our SFRD today, $\psi(z=0)$, and $R^{CC}(z\approx0)$ from the ZTF survey \citep{Perley:2020ajb} to obtain the value of $K^{CC} = f_\psi R^{CC}_{ZTF}(z\approx0)/ \psi(0)$. We introduce $f_\psi$ as a correction factor to account for the model dependency of the SFR, we find a value of $f_\psi =0.4$ to best match our model's rate to LVKs. The overall amplitude is sensitive to shifts in the prescriptions we use, such as the SFR or binary fraction and can vary by about an order of magnitude \citep{Neijssel2019}. While a more rigorous treatment would begin with the specific star formation rate, measuring it to high redshift is challenging. Pursuing that path, though compelling, lies beyond our current scope and is left for future work.

The factor of 1/2 in Equation~\ref{eq:KCC} is included to avoid double counting our systems, as we are only considering the primary mass of the binary. The overall binary fraction of stars in the galaxy is denoted by $f_{bin}$. The study by \citet{Sana:2007id} finds the intrinsic binary fraction to be $f_{\text{bin}} \sim 0.7$, while \citet{Moe2017} report that O-type stars are more commonly found in triples ($n = 3$) and quadruples ($n = 4$), with $f_{n \geq 3} = 0.73$ and $f_{\text{bin}} = 0.21$. Triples and quadruples contributing to the merger rate would consist of a closer-orbit binary pair with additional stellar companions. While we do not explicitly factor in $n \geq 3$ systems, our model accounts for a subset of their evolution. Thus, $f_{\text{bin}} \in [0.21, 0.94]$. For consistency with both studies, we adopt $f_{\text{bin}} = 0.7$ throughout.

\subsection{Volumetric Binary Black Hole Formation Rate}

In the last section we derived the per-galaxy instantaneous production rate of merging binary black holes with primary mass $M_1$ at a given redshift, $d^2N(M_\star,z)/dM_1dt$ (Equation~\ref{eq:dNdM1dt}). In this section we will extend this analysis to a volumetric population by convolving the galaxy stellar mass function (GSMF), $\phi(M_\star,z)$, which quantifies the number of galaxies per unit volume per dex in galaxy stellar mass with our binary black hole formation rate, defined as:

\begin{equation}
\dot{n}_{\rm BBH}(M\star, M_1, z)=
\phi(M_\star, z)\,
\frac{d^2N_{\rm BBH}}{dM_1\,dt}(M_\star, z)\,,
\label{eq:ndot}
\end{equation}

The quantity $\dot{n}_{\rm BBH}(M\star, M_1, z)$ represents the contribution to the volumetric BBH formation rate density per primary mass interval and per dex in galaxy stellar mass. To account for the distribution of galaxies across stellar mass and redshift, we adopt an observationally constrained star-forming GSMF. For $z < 0.2$, we use the double-Schechter fit from \citet{Baldry2012}. For $0.2 \le z \le 5.5$, we adopt the GSMF from \citet{COSMOS2020}, who fit a double-Schechter function for $z \lesssim 2.5$ and a single-Schechter function at higher redshift. To ensure smooth evolution between bins, we linearly interpolate the Schechter parameters between the bin centers. We extrapolate the redshift evolution of the Schechter parameters beyond $z = 5.5$ using a power-law dependence on $(1+z)$, calibrated to the observed trends at $z \ge 2$:

\begin{align}
    M^\ast(z) &\approx 1.42\times10^{12}\, (1+z)^{-2.376}, \\
    \phi_1(z) &\approx 7.52\times10^{-4}\, (1+z)^{-0.882}.
\end{align}

In this extrapolation, we keep $\alpha_1$ fixed at its constant value, consistent with the single-Schechter regime in \citet{COSMOS2020}. These high-redshift systems contribute minimally to the low-redshift merger rate density due to the short lifetimes of progenitor stars and the preference for shorter delay-times in our systems.

\begin{figure}
    \centering
    \includegraphics[width=0.99\linewidth]{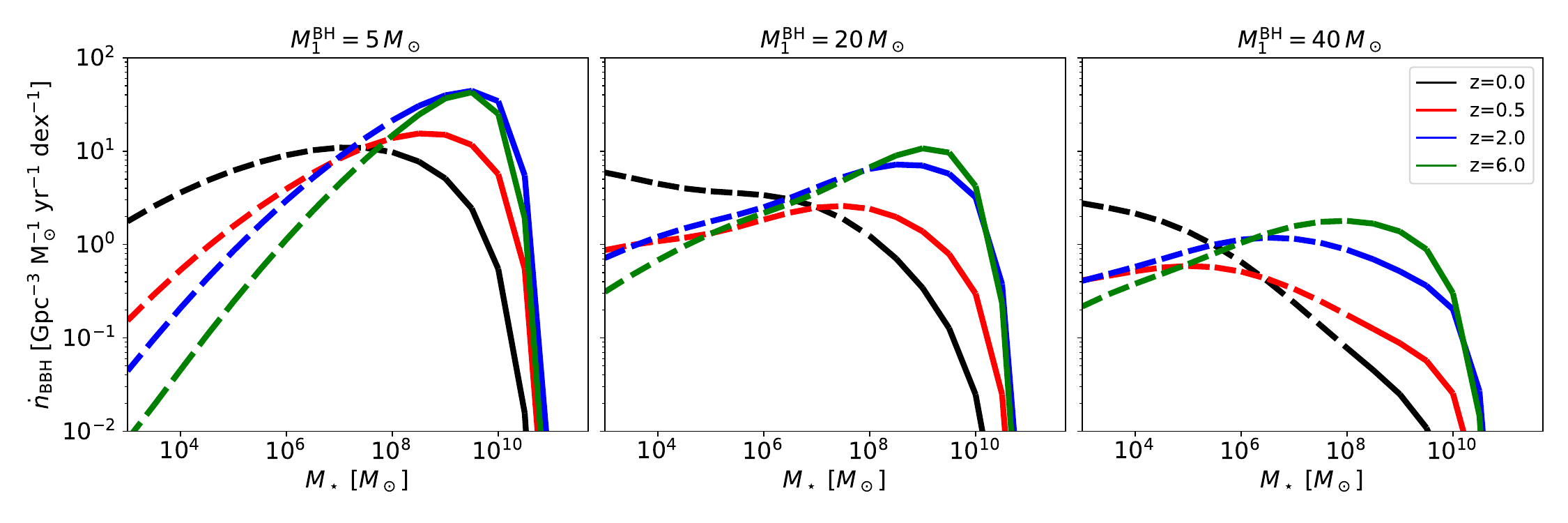}
    \caption{ Differential binary black hole (BBH) formation rate density, $\dot{n}_{\rm BBH}(M_\star,M_1,z)$, as a function of galaxy stellar mass  for three representative primary black hole masses (left to right: $M_1 = 5$, $20$, and $40\rm \,M_\odot$). Each panel shows four redshifts ($z=0$, $0.5$, $2$, and $6$), illustrating how BBH formation shifts across the galaxy population over cosmic time. Binary black hole formation migrates to lower galactic stellar masses as the Universe evolves, this migration occurs quicker for more massive BBHs. Dashed segments indicate the regime below $M_\star = 10^8\rm \,M_\odot$, where the galaxy stellar mass function is extrapolated to lower stellar masses where observational data is not yet constraining.}
    \label{fig:ndot}
\end{figure}

The results of Equation~\ref{eq:ndot} are shown in Figure~\ref{fig:ndot}. Each panel corresponds to a different primary black hole mass, illustrating how the differential BBH formation rate density, $\dot{n}_{\rm BBH}(M_\star,M_1,z)$, depends on the stellar mass of the host galaxy and on cosmic time. At early epochs near the peak of cosmic star formation ($z\simeq2$), the formation of both $5\rm \,M_\odot$ and $20\rm \,M_\odot$ black holes occurs primarily in relatively massive galaxies, whereas the highest-mass ($40\rm \,M_\odot$) black holes are already associated with lower-mass systems. By $z\simeq0.5$ ($\sim4.5$~Gyr ago), lower-mass galaxies begin dominating the production of binary black holes, with the most massive black holes forming preferentially in smaller hosts. At the present epoch, the formation of $40\rm \,M_\odot$ black holes peaks in the lowest mass galaxies, while even $5\rm \,M_\odot$ systems predominantly form in smaller dwarf galaxies. The dashed portions of each curve mark the region below $M_\star=10^8 \rm \,M_\odot$, where the galaxy stellar mass function is extrapolated beyond the range constrained by current observations; results in this regime should therefore be interpreted with caution. While \citet{chruslinka2019mnras07} show that extrapolating the GSMF to low stellar masses can lead to an overestimation of the star formation rate at high redshift, our adopted GSMF does not exhibit such an excess at early times, although it tends to overestimate the SFR at our current epoch. Overall, the figure highlights a clear shift of black hole formation toward progressively lower-mass galaxies as the Universe evolves.

\subsection{The Merger Rate Density}
\label{sec:mrdcalc}

\begin{figure}[t!]
    \centering
    \includegraphics[width=0.5\columnwidth]{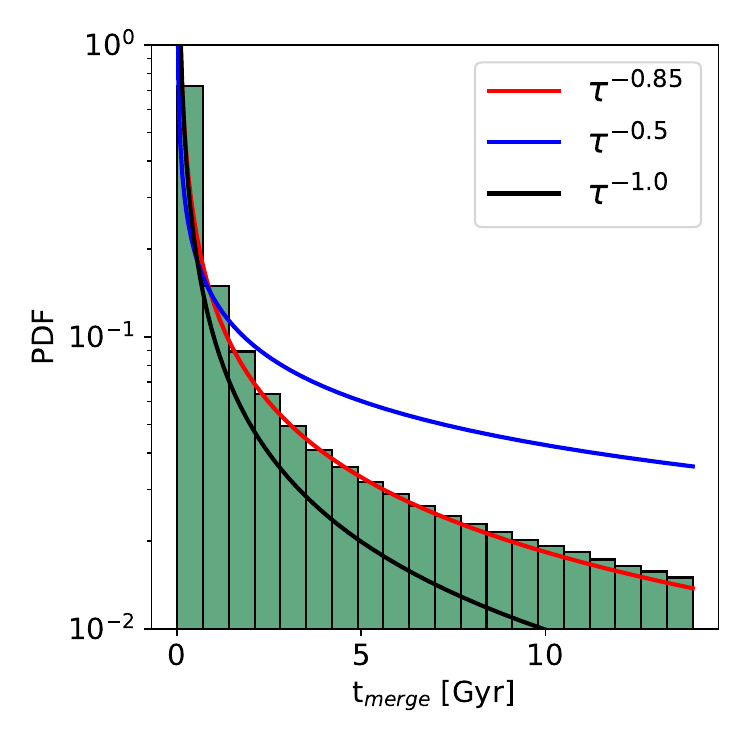}
    \caption{Power-law fit to the merger time distribution from the \texttt{SEVN} simulations. The distribution is steeply peaked at short delays, implying that the merger rate density today will be dominated by binaries formed more recently in cosmic history. We find the best fit to our data is a power-law $p(\tau) \propto \tau^{-\gamma}$ with spectral index $\gamma=0.85$. We include a steeper distribution of $\gamma=1.0$ \citep{Fishbach:2021mhp} and a flatter more extreme value of $\gamma=0.5$ for comparison.}
    \label{fig:mergertime}
\end{figure}

To obtain the merger rate density, we require the delay-time distribution (DTD), or the time between binary formation, taken to be the time the stars enter the main sequence, and black hole merger. The delay time, $\tau$, depends on orbital parameters, stellar evolution prescriptions, and the mass-loss history. We calculate the DTD, $P(\tau)$, directly from our simulations, as shown in Figure \ref{fig:mergertime}. Due to the high mass of the stars that we consider, their average lifetimes are negligible in comparison to the merger time. We fit a power-law function of the form $P(\tau) \propto \tau^{-0.85}$ to the DTD, similar to that found in \citet{Neijssel2019,Mukherjee:2021bmw,Fishbach:2021mhp, Karathanasis:2022rtr}. We will show in the following section how even modest changes to the DTD spectral index shapes the merger rate. Convolving our Equation \ref{eq:dNdM1dt} with the GSMF and the DTD we obtain the differential merger rate density, at a given time $t_0$:

\begin{equation}
    \frac{d \mathcal{R}}{d M_1} =  \int^{t_{0}}_{0}\int^{M_{max}}_{M_{min}} \phi(M_{\star},z(t_{0} - \tau ')) \frac{d^2N_{BBH}}{dM_1 \,dt}(M_\star,t_{0} - \tau ') P(\tau ') d\tau '  dM_{\star} \,.
    \label{eq:diffmrd}
\end{equation}

The limits of integration over galaxy stellar mass are set to $M_{\min} = 10^8\,\rm M_\odot$ and $M_{\max} = 10^{12}\,\rm M_\odot$, consistent with the observational range of the galaxy stellar mass function from \citet{COSMOS2020}. Although galaxies with smaller stellar masses exist, they are not well constrained observationally, and extrapolating the GSMF below this limit can lead to an overestimation of the star formation rate at low redshift. We find that extending the lower limit down to $10^3\,\rm M_\odot$ 
affects the inferred merger rate density in the next section by less than a factor of two.

\section{Results}
\label{sec:results}

\subsection{Primary Mass Spectrum}
\label{sec:bbhspectrum}

In the previous section, we detailed the calculation of the merger rate density given in Equation \ref{eq:diffmrd}. Throughout this section, we compare our results to inferred distributions based on the gravitational wave observations from the LVK collaboration and find significant agreement with their findings. 

\begin{figure*}
    \centering
    \includegraphics[width=\textwidth]{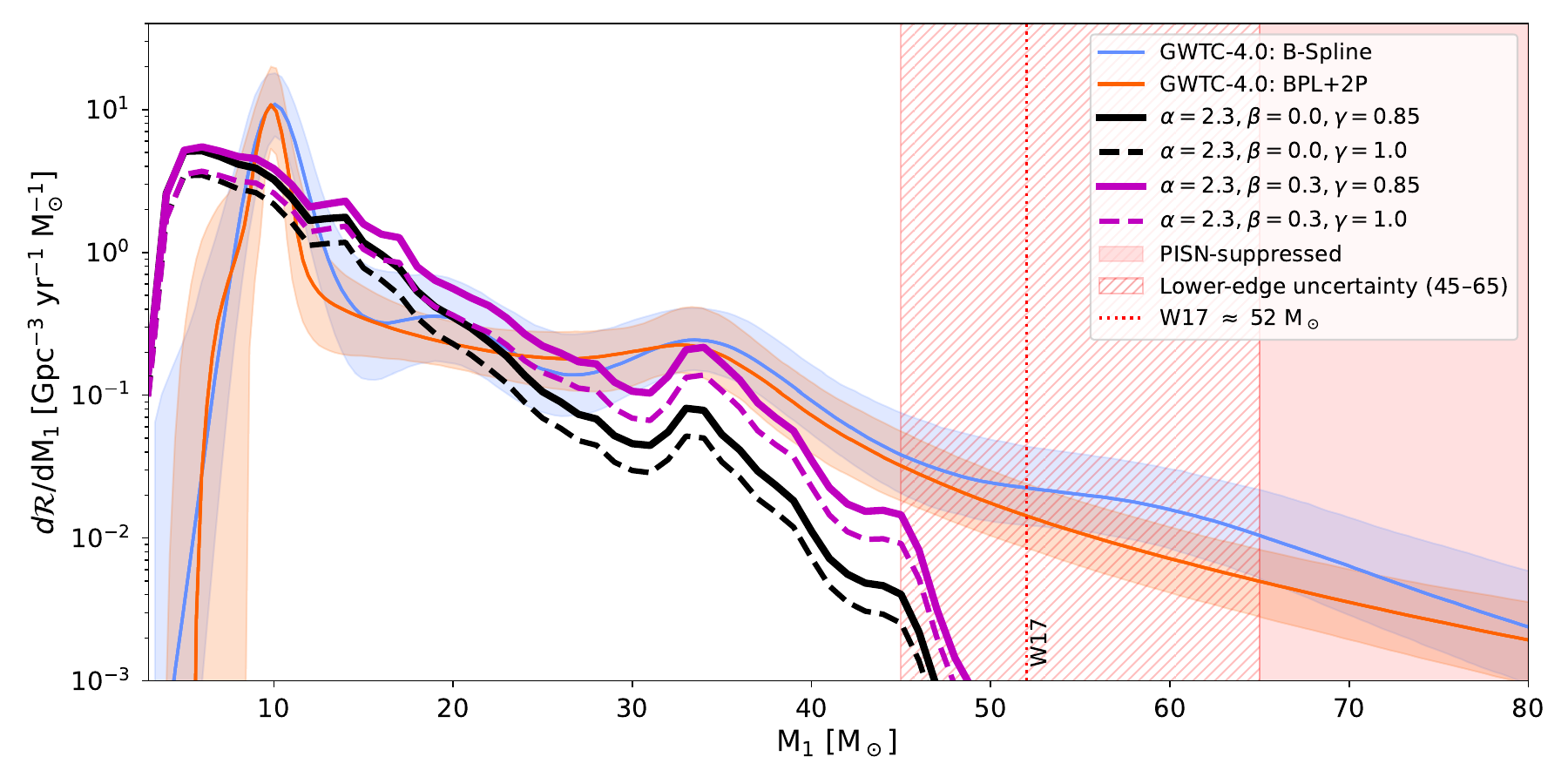}
    \caption{Comparison between the LIGO-Virgo-KAGRA \textit{Broken Power Law + Two Peaks (BPL+2P)} and \textit{B-Spline} models, shown in orange and blue, respectively, and our two IMF models: a constant-slope IMF with $(\alpha = 2.3, \, \beta = 0.0)$ and a running-slope IMF with $(\alpha = 2.3, \, \beta = 0.3)$. We also compare the impact of varying the delay-time distribution, $P(\tau)\propto \tau^{-\gamma}$, for $\gamma = 0.85$ and $1.0$. The PISN threshold is given by the red band, reflecting uncertainty in the lower boundary ($\sim45$--$65 \, \rm M_\odot$). The SEVN M20 PISN implementation follows the prescription of \citet{Woosley2017} (W17), whose threshold is also included. Our model produces very few BBH mergers within the pair-instability mass gap, consistent with standard stellar evolution theory, leading to a natural truncation around $\sim50 \,\rm M_\odot$.}
    \label{fig:drdm1}
\end{figure*}

In the most recent gravitational-wave transient catalog \citep{GWTC4pop}, the LVK Collaboration models the merger rate density and its derivatives using both parametric and non-parametric approaches. In Figure \ref{fig:drdm1}, we compare our results to two of their reference models: the \textit{Broken Power Law + Two Peaks} (BPL+2P) model, which describes a broken power-law continuum between a minimum and maximum mass with two additional Gaussian peaks at $\sim10\, \rm M_\odot$ and $\sim35\, \rm M_\odot$, and the non-parametric \textit{B-Spline} model, which represents the distribution using one-dimensional basis splines. To compare with these, we highlight two different forms for the power-law index, $\xi=-\alpha+\beta\log_{10}(M_{\rm ZAMS}/8 \rm M_\odot)$, our constant-slope ($\alpha=2.3, \beta=0.0$) which matches the fiducial Kroupa IMF \citep{Kroupa2001} and our running-slope ($\alpha=2.3, \beta=0.3$) where we introduce a non-zero $\beta$ to gradually flatten the IMF at higher masses. We find that the running-slope model exhibits greater agreement with LVK observations, suggesting that mass-dependent variations in the IMF, required mainly in dwarf galaxies, play a significant role in shaping the BBH mass distribution.

The inferred peak at $\sim35 \,\rm M_\odot$ is apparent across our models, associated with the pair-instability mechanism. The smoothing toward lower masses, present in the LVK \textit{BPL+2P} and \textit{B-Spline} models, is attributed to factors such as metallicity and binary evolution, which naturally blur the edge of the proposed lower mass gap \citep{LIGOScientific:2018jsj}. Our model exhibits a similar tapering at low masses but with the primary peak shifted to slightly lower values ($\sim6 \, \rm M_\odot$ compared to $\sim10 \,\rm M_\odot$ in the LVK results). This offset could potentially be reduced by varying the binary evolution parameters in \texttt{SEVN}; however, doing so would require running new simulations and is left to future work.

Our model demonstrates good agreement with the observed differential merger rate across much of the primary mass spectrum, including the tapering at low masses and the peaks. However, discrepancies arise at the high-mass end of the spectrum, where the model underestimates the number of binary black holes required to populate the observed tail of the merger-rate distribution. In our case, the distribution naturally tapers off around $\sim50 \, \rm M_\odot$, due to the M20 PISN model in SEVN \citet{Spera2017}, based on \citet{Woosley2017}. The true onset of the pair-instability process, however, may occur at slightly higher masses, up to $\sim65 \, \rm M_\odot$ \citep{Woosley:2021xba}. This uncertainty could contribute to the observed discrepancy at the high-mass end. Black holes above this range may arise from alternative formation channels, such as the dynamical channel, population III stars \citep{Farrell:2020zju,Kinugawa:2020xws,Liu:2020lmi}, hierarchical mergers \citep{Mapelli:2016vca,Farr2017,Mapelli2020b, Gerosa2021,Sedda:2023big,Torniamenti:2024uxl,Bruel:2025sdq}, primordial black holes \citep{DeLuca:2020sae}, or beyond the Standard Model (BSM) physics \citep{Croon:2020oga,Sakstein:2020axg,Ziegler:2020klg,Ziegler:2022apq,Croon:2023trk}.

\subsubsection{Delay-Time Distribution Effects}
\label{sec:DTD}

Included in the models in Figure~\ref{fig:drdm1} is a dependency on the delay-time distribution (DTD) power-law index. If we assume that the DTD takes the form $P(\tau) \propto \tau^{-\gamma}$ our simulations show that $\gamma=0.85$ as shown in Figure~\ref{fig:mergertime}. However, this can be influenced by many factors such as the true underlying distribution of parameters, e.g. eccentricity and orbit, as well as stellar evolution effects.

Here we briefly discuss one such possibility, that the variation in the common envelope (CE) parameters, $\alpha_{\mathrm{CE}}$ and $\lambda_{\mathrm{CE}}$, can lead to changes in the merger time and thus the DTD. We only examine one system here, and leave a full study to future work. In particular, we examine the case of a binary system with progenitor masses $M_1 = 51\,\rm M_\odot$ and $M_2 = 50\, \rm M_\odot$. In all cases where the binary survived to form black holes and merge through gravitational wave (GW) emission, the final remnant masses were fixed at $M_{1,\rm BH} = 22.7\,\rm M_\odot$ and $M_{2,\rm BH} = 22.0\, \rm M_\odot$. Depending on the CE parameters, systems either merged prematurely during the CE phase ($\order{1 \, \rm Myr}$) or survived as binary black holes with various delay times.

The CE efficiency parameter $\alpha_{\mathrm{CE}}$ controls the fraction of orbital energy used to unbind the envelope. A value greater than unity simply means that the orbital energy is not the only source removing the envelope \citep{Ivanova2013,Nandez2016, Roepke:2022icg}. Hydrodynamical studies have shown that additional energy sources, such as recombination energy, can contribute significantly to envelope ejection \cite{Nandez2016,Sand2020,Roepke:2022icg}. 

The envelope structure parameter $\lambda_{\mathrm{CE}}$ characterizes the binding energy of the envelope and depends on the stellar mass, evolutionary phase, radius, and envelope structure \citep{Heuvel1976,webbink1988}. Different prescriptions for $\lambda_{\mathrm{CE}}$ have been proposed in the literature \cite{Xu2010,Claeys2014,Klencki:2020kxd}. We refer the reader to \citet{Iorio2022} for these implementations and further details on the common envelope prescription in \texttt{SEVN}. 

The delay times and outcomes for various values of $\alpha_{\mathrm{CE}}$ and $\lambda_{\mathrm{CE}}$, varied within reason according to the \texttt{SEVN} prescriptions, are summarized in Table~\ref{tab:ce_params}. We show that altering the $\alpha_{CE}$ parameter alone, by a factor of $\sim 2$, can shift the merger time by over an order of magnitude. This is in agreement with the findings in \citet{Claeys2014}, who find that the slope of their DTD shifts when varying the common envelope efficiency. They find that for higher $\alpha_{CE}$ the DTD flattens and longer merger times are favored, while for smaller $\alpha_{CE}$ the DTD steepens, favoring shorter merger times. Decreasing $\alpha_{CE}$ leads to a lower binary black hole formation efficiency, potentially lowering the overall merger rate. While we do not quantify the full impact of such effects on the DTD in this work, we illustrate how sensitive the merger timescale is to even modest changes in stellar evolution assumptions, spanning outcomes from prompt stellar mergers to binaries with merger times exceeding the Hubble time by orders of magnitude. A comprehensive exploration of these effects is left for future study.

\begin{table}[htbp]
    \centering
    \begin{tabular}{c c l}
        \toprule
        Parameter & Delay Time [Myr] & Outcome \\
        \midrule
        \(\alpha_{\mathrm{CE}} = 5\)   & 234.4       & BBH merger (GW) \\
        \(\alpha_{\mathrm{CE}} = 4\)   & 111.9       & BBH merger (GW) \\
        \(\alpha_{\mathrm{CE}} = 3\)   & 52.1        & BBH merger (GW) \\
        \(\alpha_{\mathrm{CE}} = 2\)   & 15.7        & BBH merger (GW) \\
        \(\alpha_{\mathrm{CE}} = 1\)   & --       & Stellar merger (CE) \\
        \(\alpha_{\mathrm{CE}} = 0.1\) & --        & Stellar merger (CE) \\
        \midrule
        \(\lambda_{\mathrm{CE}} = -1\) \citep{Claeys2014}   & 52.1        & BBH merger (GW) \\
        \(\lambda_{\mathrm{CE}} = -4\) \citep{Klencki:2020kxd}   & 9.4         & BBH merger (GW) \\
        \(\lambda_{\mathrm{CE}} = -5\) \citep{Xu2010}  & 25.0        & BBH merger (GW) \\
        \(\lambda_{\mathrm{CE}} = -12\) \citep{Claeys2014}  & \(>10^6\)   & No merger (BBH) \\
        \midrule
        \(\alpha_{\mathrm{CE}} = 1, \lambda_{\mathrm{CE}} = -12\) & 444.6 & BBH merger (GW) \\
        \(\alpha_{\mathrm{CE}} = 5, \lambda_{\mathrm{CE}} = -5\)  & 104.4 & BBH merger (GW) \\
        \bottomrule
    \end{tabular}
    \caption{Delay times and outcomes for a binary with $M_{p,1} = 51\,\rm M_\odot$ and $M_{p,2} = 50\,\rm M_\odot$, evolved under different common envelope (CE) parameter choices in \texttt{SEVN}. All systems that survive CE and undergo GW-driven inspiral result in BBH mergers with $M_{1,\mathrm{BH}} = 22.7\,\rm M_\odot$ and $M_{2,\mathrm{BH}} = 22.0\,\rm M_\odot$. The $\alpha$-formalism for CE evolution in \texttt{SEVN} is adopted from \citet{Hurley:2000pk}. For the CE binding energy parameter $\lambda_{\mathrm{CE}}$, we follow \texttt{SEVN}’s convention of using integer flags to select different parameterizations and denote which study they are from in the Table. The default parameters in \texttt{SEVN} are $\alpha_{\mathrm{CE}} = 3$ and $\lambda_{\mathrm{CE}} = -1$. In each row, we indicate which parameter is varied while the other is held fixed at its default value, except in the last two rows, where both parameters are modified simultaneously. } 
    \label{tab:ce_params}
\end{table}

\subsubsection{Dynamical Single Black Holes}

An intriguing avenue to explore the population of mergers within the upper mass gap lies in the role of single black holes. Although the fraction of massive stars born as singles is relatively low (around $10\%$), these single black holes evolve in isolation from ZAMS to BH and are not subject to the same efficiency factor constraints as shown in Figure \ref{fig:eff_v_Z}. As a result, if they retain their hydrogen envelopes, they can form more massive black holes, even extending into the mass gap.  

Once formed, single black holes can enter binaries through dynamical capture \citep{Mapelli:2016vca,Gerosa2021}, a process that depends on several factors, including the cluster density, stellar interaction rates, and black hole retention within these environments. After formation, black holes tend to migrate toward the cluster core due to dynamical relaxation \citep{Portegies2000}. Closer to the cluster core, the more frequent dynamical encounters can result in the formation of binaries or higher-order systems. While many black holes will experience multiple interactions or even ejection from the cluster, some may successfully become part of a binary system capable of merging within a Hubble time.

We sketch a simplified picture to estimate the efficiency of this process, $\eta$, which we define as the probability for a single black hole to capture a companion (i.e., form a binary) and merge within the age of the universe. We can break up this efficiency into various factors to get a better idea;  $\eta \sim f_{\text{sse}} \times f_\tau \times f_{\text{dyn}} \times f_{\text{GC}} \times f_{\text{primary}}$. Here, $f_{\text{sse}}$ represents the fraction of massive stars born as singles, $f_\tau$ is the probability for dynamically formed binaries to merge within a Hubble time, $f_{\text{dyn}}$ accounts for the fraction of dynamically interacting black holes that form binaries instead of being ejected that is likely dependent on metallicity~\citep{Kumamoto2020}, $f_{\text{GC}}$ quantifies the fraction of black holes residing in dense stellar environments like globular clusters (which we expect to be dependent on the stellar mass of the galaxy and perhaps its metallicity), and $f_{\text{primary}}$ reflects the likelihood that the single black hole becomes the primary in the binary. These various factors depend on the mass of the single black hole, which would require detailed simulations to explore.
 
To explore the possibility of binaries formed dynamically contributing to the upper mass gap, we decompose the total differential merger rate into two components:
\begin{equation}
    \frac{dR}{dM_1} = \frac{dR_{\text{BBH}}}{dM_1} + \bar{\eta}(M_1) \frac{dR_{\text{SBH}}}{dM_1} \,,
\end{equation}
where $\frac{dR_{\text{BBH}}}{dM_1}$ corresponds to the contribution from black holes formed in isolated binaries, while $\frac{dR_{\text{SBH}}}{dM_1}$ accounts for the contribution from single black holes and $\bar{\eta}$ is the efficiency averaged over galaxy masses and metallicities. Given our discussion above, we should build in the efficiency calculation before averaging over the properties of galaxies but we are only interested in an order of magnitude answer here. 
If the black hole mass dependence of $\bar{\eta}$ is neglected, we can do a quick estimate of the efficiency needed to populate black holes in the mass gap. 
We estimate that, for an efficiency $\bar{\eta} \sim 10^{-4}$, the contribution from dynamically formed binaries can extend the black hole mass spectrum up to about $60 \, \rm M_\odot$, about $\sim 10 \,\rm M_\odot$ further into the mass gap than our model. 

It is also worth keeping in mind that the lower mass when pair instability kicks in is not well known. The effects discussed in \citet{Woosley:2021xba} suggest that the pair instability could shift to higher masses ($\sim 65$–$70 \, \rm M_\odot$) and this would offer an additional way to populate the mass gap. It would be interesting to include these effects along with the mass loss in binary systems to explore this further. Together, these considerations present pathways to populate the upper-mass gap that deserve further study.

\begin{figure*}
    \centering
    \includegraphics[width=\textwidth]{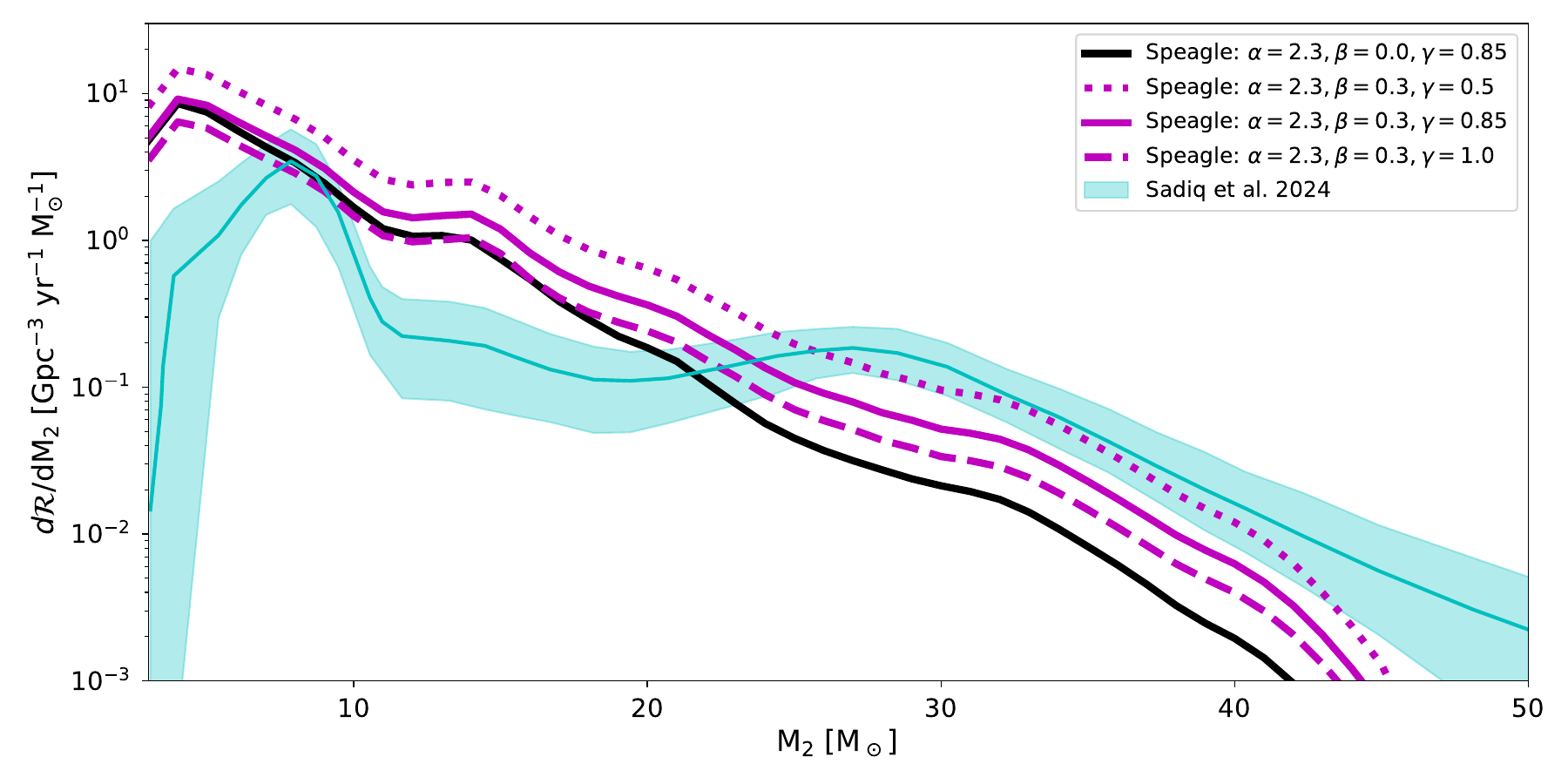}
    \caption{Differential merger rate distribution as a function of secondary black hole mass, $dR/dM_2$. We show both constant-slope ($\alpha = 2.3$, $\beta = 0.0$) and running-slope ($\alpha = 2.3$, $\beta = 0.3$) IMF models, where the running-slope model is evaluated under varying delay-time power-law indices. For comparison, we include the inferred distribution from \citet{Sadiq:2023zee}, shown as a cyan band. While our results are broadly consistent, our predicted shape follows a closer approximation to a single power-law with small bump-like features, rather than the more pronounced structure in their inferred distribution. This is particularly useful to compare against future LVK results.}
    \label{fig:dRdM2}
\end{figure*}

\subsection{Secondary Mass Spectrum}
\label{sec:secondaryspectrum}

With over 150 confirmed BBH events detected up to O4a \citep{GWTC4pop}, the LVK dataset offers growing potential to probe the component mass spectra of merging binary black holes. While the primary mass distribution has received significant attention, the secondary mass spectrum remains comparatively underexplored, though recent work has begun to address this gap \citep{Sadiq:2023zee,Farah:2023swu}. While the LVK \textit{FullPop-4.0} model provides posterior distributions for both component masses, the published secondary-mass spectrum is limited to $m_2 \lesssim 15\, \rm M_\odot$ and represents the full compact-object population, including both neutron stars and black holes. As such, it does not isolate the secondary-mass distribution for binary black holes alone. Since a BBH-specific secondary spectrum has not yet been released, we compare our results to the inference-based study of \citet{Sadiq:2023zee}. Unlike the inference-based approaches of \citet{Sadiq:2023zee,Farah:2023swu}, our framework models the astrophysical population by combining galaxy and cosmological evolution with stellar population synthesis, allowing for a self-consistent exploration of the origin and properties of the secondary mass distribution.

Starting from Equation~\ref{eq:dNdM1dt}, and omitting the integration over mass ratio $q $, we obtain the differential distribution $\frac{d^2N}{dM_1\,dq\,dt}(M_\star, z) $. To express this in terms of the secondary mass $M_2 = q M_1 $, we perform a change of variables in phase space, applying the appropriate Jacobian transformation:

\begin{equation}
    \frac{d^2N}{dM_2\,dt}(M_\star, z) = \int_{M_1 = M_2}^{80 \, \rm M_\odot} \left| \frac{dq}{dM_2} \right| \frac{d^3N}{dM_1\,dq\,dt}(M_\star, z) \, dM_1 \,.
\end{equation}

This transformation requires evaluating $ \frac{d^2N}{dM_1\,dq\,dt} $ outside our simulated grid for values of $ (M_1, q) $, since $ M_2/M_1 $ will generally not lie exactly on the original simulation grid. To handle this, we use a bivariate interpolation across the simulated grid of $ M_1 $ and $ q $ to estimate the distribution at arbitrary values. We then follow the steps in Section \ref{sec:mrdcalc} to obtain the secondary mass spectrum $d\mathcal{R}/dM_2$ at  $ z = 0.2 $ which we show in Figure~\ref{fig:dRdM2}.

Compared to the primary mass distribution, the secondary mass spectrum features a larger number of low-mass events, $\lesssim 8 \, \rm M_\odot$, but the rate declines more steeply at higher masses. Except for this overabundance at low masses the secondary spectrum remains below the primary mass spectrum in terms of overall events. The peak which is present in the primary spectrum at $\sim 35 \, \rm M_\odot$ is flattened to a small bump like feature and shifted to slightly lower masses in the secondary spectrum. Within the peak mass range, $\sim 30-40 \, \rm M_\odot$, the spectra are dominated by low metallicity systems. Contributions from metallicities in the range $10^{-3} \leq Z <10^{-4}$ exhibit a mass ratio near unity, thus both the primary and secondary spectra have contributions here. However, the primary mass spectrum is unique in that it obtains a boost of systems in the intermediate metallicity regime between $10^{-2} < Z <10^{-3}$, yet the mass ratio strays from unity and these systems in the case of the secondary mass spectrum get spread out to lower masses weakening the peak like feature.

Similar to the primary mass spectrum in Figure~\ref{fig:drdm1} we explore the effect of varying the delay-time distribution (DTD), $ P(\tau) \propto \tau^{-\gamma} $, using our fit value of $\gamma=0.85$, a steeper distribution with $\gamma=1.0$, and a flatter value of $ \gamma = 0.5 $ leading to longer mergers. The effect of changing the DTD is effectively an amplitude shift. The impact or DTD variations on the spectral shape is negligible. On the other hand, the variation in the IMF can lead to order of magnitude differences at the high-end tail of the mass distribution when considering our top-heavy model (magenta lines).

For comparison, we include the results from \citet{Sadiq:2023zee}, who infer the secondary mass distribution directly from LVK data. Their results exhibit more pronounced features, including a stronger bump near $ \sim 30\,M_\odot $. In fact \citet{Farah:2023swu} show that the high mass peak can be stronger in the case of the secondary mass spectrum. While our model does not reproduce this structure, the overall slope of our predicted distribution aligns more closely with a single power-law, featuring only mild bump-like deviations. If such a feature is confirmed with higher significance in O4, it may indicate the contribution of a distinct formation channel, such as dynamical assembly in dense stellar environments, or possibly hint at new physics beyond standard stellar evolution models. As with the primary mass spectrum, we observe a sharp decline in the secondary mass spectrum above $ \sim 50\, \rm M_\odot $, consistent with the expectations from our prescription of pair-instability supernovae.

\begin{figure}[t!]
    \centering
    \includegraphics[width=0.5\columnwidth]{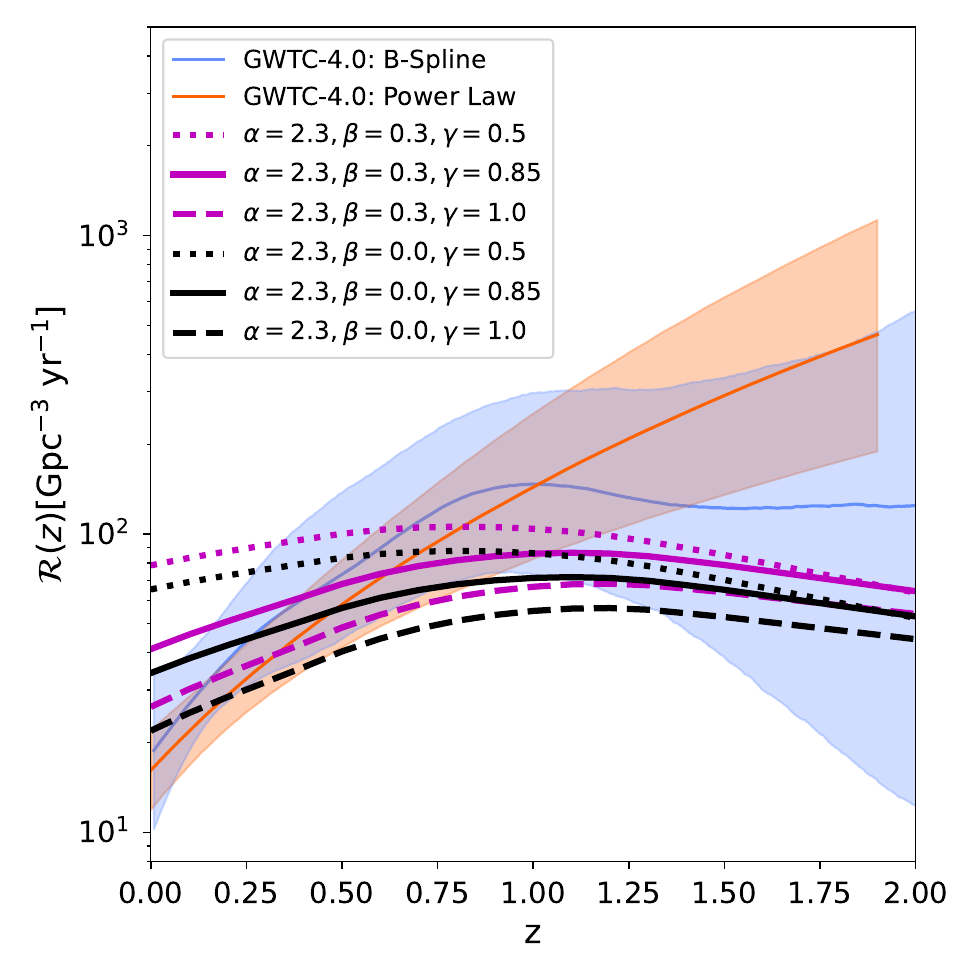}
    \caption{
        Total binary black hole merger rate density as a function of redshift for our constant- and running-slope IMF models ($\alpha = 2.3$, $\beta = 0.0$ and \(0.7\), respectively), evaluated under three different delay-time distribution power-law indices: $\gamma = 0.5$, $0.85$, and $1.0$. The shaded bands represents the 90\% credible interval inferred from LVK observations. Models with longer delay-times ($\gamma = 0.85$ and $1.0$) show better agreement with the inferred redshift evolution, while the $\gamma = 0.5$ model overpredicts the low-redshift rate. 
    }
    \label{fig:drdz}
\end{figure}

\subsection{Redshift Dependence}
\label{sec:mrd_redshift}

To obtain the total merger rate at a given redshift, we first determine $d\mathcal{R}/dM_1$ at the corresponding cosmic time, $t_0 = t(z)$, using Equation \ref{eq:diffmrd}. We then integrate over the black hole primary mass to find $\mathcal{R}(z)$:

\begin{equation}
    \mathcal{R}(z) = \int \frac{d\mathcal{R}}{dM_1}(z) dM_1 \,.
\end{equation}

The results of the redshift evolution of the total merger rate density are given in Figure~\ref{fig:drdz} for both our constant- and running-slope IMF models, each evaluated at three different delay-time distribution (DTD) power-law indices: $\gamma = 0.5$, $0.85$, and $1.0$. These values bracket the distribution favored by our simulations ($\gamma = 0.85$) and explore both steeper and shallower scenarios, corresponding to shorter and longer delay-times. We choose $\gamma=1.0$ as this is the value that is inferred from the LVK data in \citet{Fishbach:2021mhp}. The value of $\gamma =0.5$ is chosen to highlight how the flattening of the DTD affects the redshift evolution of the merger rate, while extreme is not completely ruled out \citep{Mukherjee:2021bmw,Karathanasis:2022rtr}.

The $\gamma=0.5$ case produces a flatter evolution, over predicting the rate at low $z$ and is overall a poor fit to the inferred spectrum. Models with shorter delay-times ($\gamma = 0.85$ and $1.0$) show better agreement with the LVK-inferred trend remaining mostly in the 90\% credible interval for their \textit{B-Spline} model, whle also aligning with their peak location. The peak at lower redshift than the power-law model can be attributed in part due to our adopted star-formation rate density, which peaks at z $\approx$ 1.5 (Appendix~\ref{app:mssfr}), and is further affected by the convolution with the delay-time distribution and the effects of stellar-evolution prescriptions. For instance, \citet{Boesky:2024msm,Boesky:2024a} show that changes to common envelope efficiency, stable mass transfer efficiency, and supernova natal kick dispersions can shift the peak redshift of the total merger rate. There are also uncertainties surrounding the galactic and cosmological inputs which can impact the merger rate \citep{Chruslinska:2018hrb,Neijssel2019,chruslinka2021ifs,vanSon:2022ylf}.

The differences between the constant- and running-slope models that we present are minimal. The similarity between the two models, for each $\gamma$, is because the total merger rate is primarily dominated by low-mass systems. The running-slope IMF is designed to enhance the high-mass end of the black hole mass spectrum, but has limited impact on the total rate due to the suppression of the high-mass tail. Lastly, we only present results on the isolated channel, it is feasible that the dynamical channel may supplement our total merger rate density, at higher redshifts, improving agreement with LVK. 

\begin{figure}[t!]
    \centering
    \includegraphics[width=0.48\linewidth]{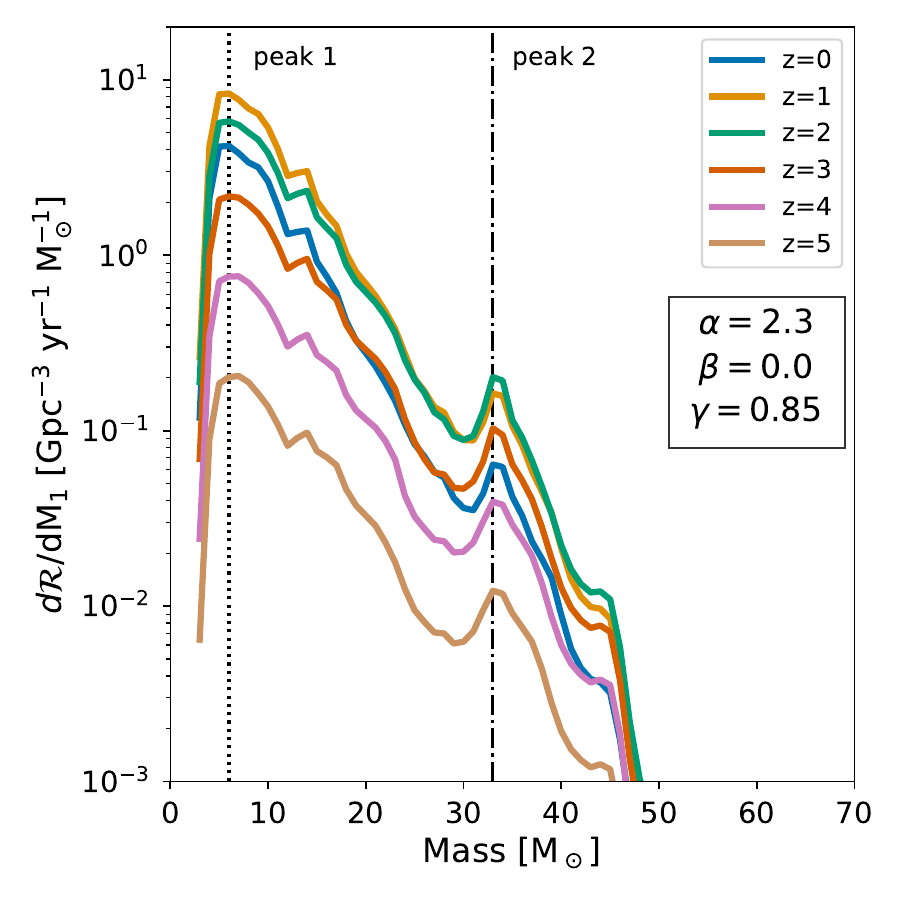}
    \hfill
    \includegraphics[width=0.48\linewidth]{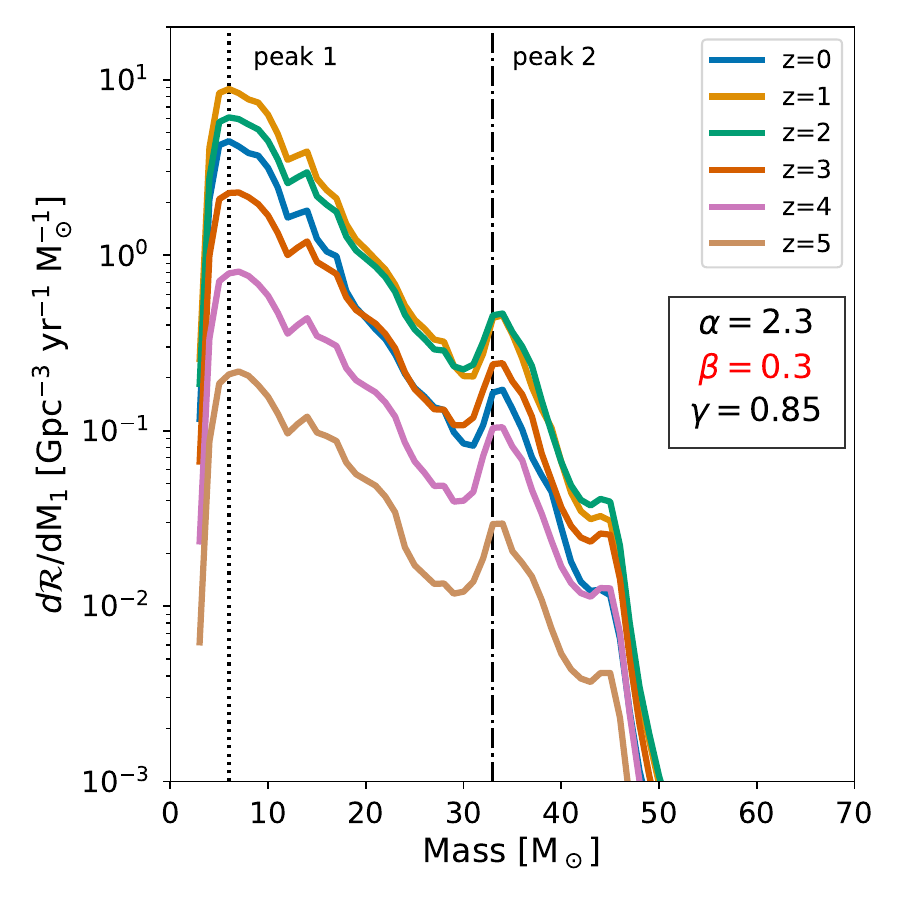}
    \caption{Binary black hole primary mass spectrum across cosmic time. Included are results from our constant-slope model (left) and running-slope model (right). The location of the features of the mass spectrum are largely robust against time as can be shown via the vertical lines. However, the peak's relative abundances does vary with time.  This can be understood via the redshift evolving metallicity-mass relation \citep{Ma:2015ota} where high-metallicity low-mass black holes become increasingly abundant relative to low-metallicity high-mass black holes as time evolves.}
    \label{fig:drdm1_z}
\end{figure}

We further explore redshift evolution in Figure~\ref{fig:drdm1_z}, which shows how the primary mass spectrum, $d\mathcal{R}/dM_1$, evolves over cosmic time for both our constant-slope (left panel) and running-slope (right panel) models. The location of the peaks in $d\mathcal{R}/dM_1$ remains robust across redshift, which is to be expected as they arise from the same underlying physical processes. The primary change we find is in the slope between the low-mass and high-mass peaks: as redshift increases (moving to earlier times), the slope becomes flatter. This evolution is more pronounced in our constant-slope model (fiducial Kroupa IMF) than in the running-slope model ($\beta=0.3$) which is constructed to increase the relative amount of low-metallicity high-mass black holes, leading to a growing contribution from high-metallicity low-mass black holes relative to low-metallicity high-mass black holes.

These trends, the static location of the peaks and slowly varying slope, are consistent with the observations of LVK data \citep{Fishbach:2021yvy,GWTC3pop,lallemam25}. The study by \citet{lallemam25} show that location of the $\sim 35\, \rm M_\odot$ peak is constant across redshift observable by LVK and that the slope of the primary mass spectrum only slightly varies over this time, consistent with our findings in Figure \ref{fig:drdm1_z}. On the other hand, \citet{Rinaldi:2023bbd} report the overall peak and spectral shape of their primary mass distribution shifts substantially with redshift attributing this to contributions from two distinct BBH populations. Since our analysis is restricted to the isolated channel, other formation pathways could, in principle, shift the primary mass spectrum at earlier redshifts. In contrast, the evolution in our models is insufficient to produce the pronounced spectral changes reported in their work, the distribution would remain dominated by the low-mass peak at $\sim 8\, \rm M_\odot$ across redshift. Similarly, it was shown that if the PISN mass scale was allowed to vary with metallicity, the peak at $\sim 35\, \rm M_\odot$ would shift with redshift \citep{Karathanasis:2022rtr}. While this is an interesting approach, our models use a fixed PISN prescription at each metallicity, so we do not observe such a shift, inline with LVK findings.

\begin{figure}[t!]
    \centering
    \includegraphics[width=0.5\columnwidth]{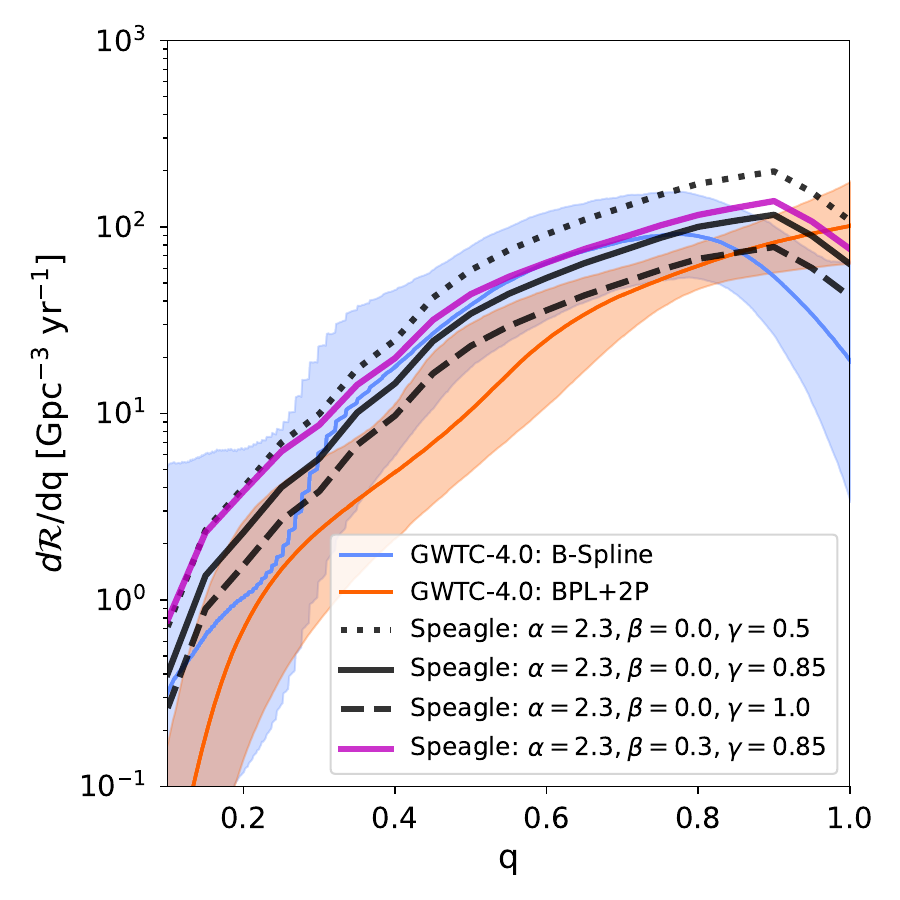}
    \caption{Comparison of the mass-ratio distribution for our constant- and running-slope models with the 90\% credibility intervals from the LVK \textit{BPL+2P} (orange) and \textit{B-Spline} (blue) models \citep{GWTC4pop}. Both our models and the LVK non-parametric \textit{B-Spline} analysis exhibit a rise toward equal-mass binaries with a turnover near $q\sim0.8-0.9$, followed by a decline at higher $q$ values.}
    \label{fig:drdq}
\end{figure}

\subsection{Mass Ratio}
\label{sec:mrdq}
We conclude our comparison with the LVK's binary black hole observations by analyzing how the merger rate depends on the mass ratio of the black holes, $q_{BH} = \frac{M_2}{M_1}$, as this has been inferred by LVK. To achieve this, we use Equation \ref{eq:dNdM1dt} integrating over the full range of primary masses, $M_1$, rather than mass ratio $q$:

\begin{equation}
    \frac{d^2N}{dq \, dt} (M_\star ,z) = \frac{S(z,M_\star)}{\langle M_{\rm ZAMS}\rangle(M_\star,z)} \, \int^{M_u}_{M_l} \, dM_1 \int^{Z_{u}}_{Z_{l}} dZ \, \rho (Z | M_\star,z) \int_{8 M_\odot}^{200 M_\odot} dM_{\rm ZAMS} \, p(M_{\rm ZAMS}| z, M_\star) \, p(M_1, q|Z,M_{\rm ZAMS}) \,,
\label{eq:dNbbhdq}
\end{equation}

where, $ p(M_1, q \mid M_{\rm ZAMS}, Z) $ represents the joint probability of forming a black hole with primary mass $M_1$ and mass ratio $q$, , as described in Section \ref{sec:sims}. The limits for the metallicity integral match our simulation limits $Z_{u}=1.5\times 10^{-6}$ and  $Z_{u}=3.2\times 10^{-2}$, and the limits for the primary black hole mass are taken from $M_l=3 \, M_\odot$ to $M_u=80 \, M_\odot$. Building on Equation \ref{eq:dNbbhdq}, we integrate over the delay-time distribution (DTD) and galaxy stellar mass function (GSMF) to obtain:

\begin{equation}
    \frac{d \mathcal{R}}{dq} =  \int^{t_{0}}_{0}\int^{M_{max}=10^{12} \, \rm M_\odot}_{M_{min}=10^8 \, \rm M_\odot} \phi(M_{\star},z(t_{0} - \tau ')) \frac{d^2N_{BBH}}{dq \,dt}(M_\star,t_{0} - \tau ') P(\tau ') d\tau '  dM_{\star}\,.
\label{eq:drdq}
\end{equation}

Our results are generally consistent with the LVK mass ratio spectrum at z=0.2, following the same overall trend of increasing merger rate density with q as shown as shown in Figure~\ref{fig:drdq}. We show the 90\% C.I. from  from LVK's \textit{BPL+2P} and \textit{B-Spline} models. Our models exhibit the best agreement with the non-parametric \textit{B-Spline} model. We include our constant- and running-slope models, with varying delay-time distribution power-law indices. Both reproduce the overall rise toward equal-mass binaries, though our distributions exhibit a turnover at slightly higher mass ratios, $q \sim 0.9$ rather than $q \sim 0.8$, inline with previous inferences \citet{Edelman:2022ydv,Farah:2023swu,Rinaldi:2025emt}. This feature has been linked to stable Roche-lobe overflow systems \citep{vanSon:2021zpk} and variations in mass-transfer efficiency in \texttt{SEVN} could shift this to lower mass ratio. With the completion of the O4 observing run and an expanded catalog of detections, the shape of the high-$q$ regime will be better constrained, providing an important test of these models.

\begin{figure}
    \centering
    \includegraphics[width=0.8\columnwidth]{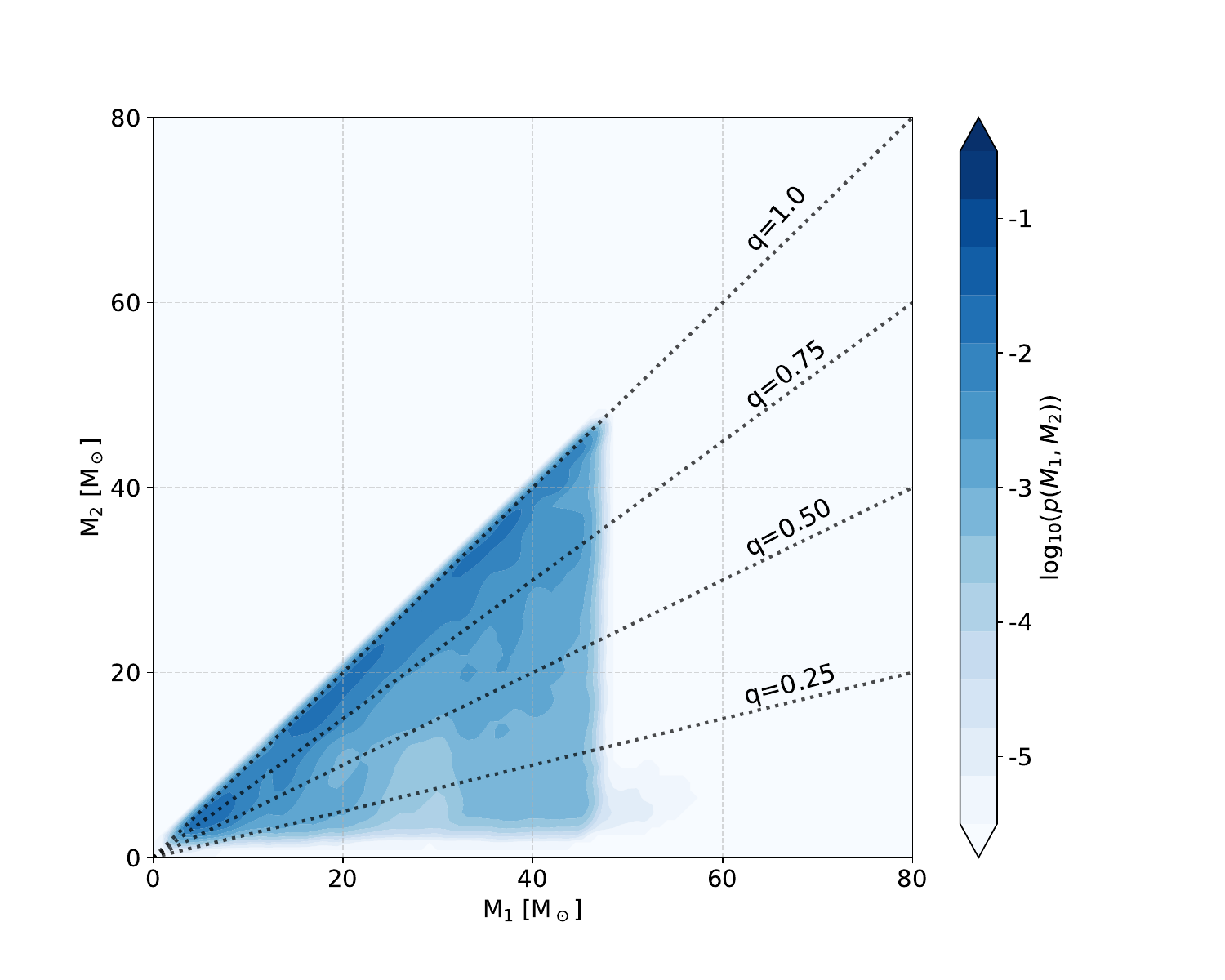}
    \caption{Joint distribution of primary and secondary black hole masses ($M_1$--$M_2$) for binaries that merge within a Hubble time in our simulations. The color scale indicates the normalized density $\log_{10}(\rho(M_1,M_2))$, and black dotted lines mark contours of constant mass ratio $q = M_2/M_1$. Binary systems cluster toward $q \to 1$, reflecting the tendency for near equal-mass mergers. Overdense regions appear near $M_1 \sim 10\,M_\odot$, $20\,M_\odot$, and $35\,M_\odot$, consistent with the primary mass features inferred by LVK.}
    \label{fig:m1m2contour}
\end{figure}

To connect the primary and secondary mass spectra to the mass ratio distribution, Figure \ref{fig:m1m2contour} shows the joint distribution of $M_1$ and $M_2$ for systems in our simulations that form BBHs merging within a Hubble time. This can be compared to the inferred two-dimensional underlying mass distributions from LVK data in \citet{Farah2023} (their Figure 2) and \citet{Callister:2024cdx} (their Figure 7). While our simulation outputs do not represent the full astrophysical population, since they have not yet been convolved with our galactic and cosmological relations, they are illustrative for understanding how features in the simulated population propagate into the merger rate.

The simulated $M_1$--$M_2$ distribution shows over-densities near $10 \, \rm M_\odot$, $20 \,M_\odot$, and $35 \, \rm M_\odot$ in the primary mass spectrum. The lower-mass excesses appear in both $d\mathcal{R}/dM_1$ and $d\mathcal{R}/dM_2$, whereas the $20\, \rm M_\odot$ excess is largely washed out, though their is a small peak remnant in the primary spectrum. The $35\, \rm M_\odot$ excess is observable in $d\mathcal{R}/dM_1$ but suppressed in $d\mathcal{R}/dM_2$, as discussed in Section~\ref{sec:secondaryspectrum}. Furthermore,  the $M_1$--$M_2$ space reveals how systems cluster at various $q$ values. In particular, our simulations show a steady rise in the number of systems with increasing mass ratio up to $q=1$. However, as shown in Figure~\ref{fig:drdq}, the actual merger rate peaks at slightly smaller values of $q\sim 0.9$, underscoring how the raw simulation outputs are reshaped into an astrophysical population once our galactic and cosmological inputs are applied.

\section{Conclusion}
\label{sec:conclusions}

Our approach, grounded in observed galactic and cosmological relations, utilizes the \texttt{SEVN} stellar evolution code to construct a model of the binary black hole (BBH) population, which are found to merge within the age of the universe. We folded in the redshift dependent relations for the metallicity mass relation (MZR), galaxy stellar mass function (GSMF), and star-forming main sequence to compute the binary black hole merger rate density. Our results show good agreement with the LVK observations, in particular with their \textit{B-Spline} model, across the total and differential merger rates with respect to primary mass, secondary mass, and mass ratio. This work will help create better model predictions for comparison to the additional BBH mergers ($\gtrsim 200$) from observing run 4 \citep{KAGRA:2013rdx,Kiendrebeogo:2023hzf, Broekgaarden:2023rta}.

There are significant uncertainties in the various elements of our calculation, including the star formation history, the initial mass function, and the physical processes in stellar evolution theory, such as mixing, wind mass loss, and common envelope phase as modeled by the \texttt{SEVN} code~\citep{Iorio2022}. While these factors introduce variability, they also present opportunities to gain insight. Studies to systematically quantify the impact of these uncertainties have been made by \citet{Chruslinska:2018hrb,Neijssel2019, Spera2019,Tang:2019qhn,Santoliquido:2020bry,Broekgaarden:2021efa,Briel:2021bpb,Iorio2022,vanSon:2022ylf}, who explored the effects of uncertain stellar evolution and galactic and cosmological relations across cosmic time on binary compact object merger rates. These uncertainties combined can alter the merger rate upwards of an order of magnitude \citep{Neijssel2019}. Our aims in this paper are to elucidate how the various elements of the calculation come together to produce the distribution of binary black holes in a simple way, and to highlight the importance of the metallicity distribution in galaxies and other key features summarized below. We show that it is difficult to match the LVK inferred distribution through the isolated channel alone and what is needed of star formation and stellar physics to get close. Our work further shows the need for a population arising from the dynamical channel, especially at higher masses.

The predicted merger rate as a function of redshift is consistent with observations, as shown in Figure \ref{fig:drdz}, where our model mostly lies within the 90\% confidence interval across cosmic time. This is encouraging given the simplicity of the model and the reasonable value for the average core collapse supernova rate used to normalize the merger rate. Our predicted rate is robust to redshift-dependent variations in the GSMF and MZR, as the most significant uncertainties in these quantities arise at higher redshifts. This insensitivity stems from the relatively short delay-time distribution (DTD) in our model, with most mergers occurring within 1 Gyr (Figure~\ref{fig:mergertime}) of BBH formation. As a result, the majority of systems contributing to the observed LVK merger are expected to originate from lower redshifts, where the GSMF and MZR are better constrained and less variable.

A more significant source of uncertainty lies in the form of the DTD itself. We adopt a power-law parameterization, $P(\tau) \propto \tau^{-\gamma}$, though the precise value of the spectral index $\gamma$ remains unconstrained. While \citet{Fishbach:2021mhp} demonstrate that the LVK data are consistent with $\gamma = 1.0$, our simulations favor a slightly shallower distribution with $\gamma = 0.85$. We also include a more extreme value of $\gamma=0.5$ in our calculations which lies just outside the 90\% credibility interval from \citet{Mukherjee:2021bmw} ($\sim 1.6\sigma$ for a Gaussian posterior). We explore the impact of varying $\gamma$ and find that steeper values (e.g., 0.85 and 1.0) yield better agreement between our model and the LVK-inferred merger rate evolution. That our total merger rate peaks at smaller redshifts than the inferred distribution, and the star formation rate (see Appendix~\ref{app:mssfr}), stems from our choice of GSMF and star-forming main sequence, the convolution with the DTD, and our choice of stellar prescriptions which have been shown to shift the peak of the total merger rate \citep{Boesky:2024msm,Boesky:2024a}.

Furthermore, comparing our model against the inferred differential merger rates yields good agreement. The black hole mass spectrum, $d\mathcal{R} /dM_1$, has several features rich with information \citep{GWTC3pop,Tiwari:2021yvr,Farah2023}. The smooth tapering at low masses is apparent, driven by binary effects and the decreasing probability of lower-mass black holes being the primary in a BBH system. We obtain a small peak at $\sim 20 \, M_\odot$, consistent with the LVK's \textit{B-Spline} model. However, the observed abundance of black holes in the $30-40 \, M_\odot$ range presents a challenge for the model with a constant power-law stellar initial mass function.

We explored the possibility that the average IMF is top-heavy by allowing the power-law index of the IMF to vary with mass. We found that a running-slope model ($-2.3 + \beta \log_{10}(M_{\rm ZAMS}/8 \rm M_\odot)$) is a better fit to the inferred primary mass distribution, with the best results for $\beta=0.3$. Most high-mass binary black holes in our model arise in dwarf galaxies, which are metal-poor, and hence this result is essentially arguing for a more top-heavy IMF in dwarf galaxies. Furthermore, \citet{Tanikawa:2020abs,Tanikawa:2021qqi} find that coupling their top-heavy IMF with the assumption of inefficient convective overshooting can produce mass gap BHs. These results are sensitive to the evolutionary model they choose, for instance utilizing the so-called L model from \citet{Yoshida2019}, the progenitors of mass gap BHs no longer retain their hydrogen envelopes and are not formed. Similar results have been found using SEVN where \citet{Iorio2022} show that varying the core overshooting parameter and the pair instability model can nearly double the final maximum mass of a black hole. If the results in \citep{Tanikawa:2020abs,Tanikawa:2021qqi} hold for higher metallicity systems, like those in this work, then it would be possible for the isolated channel to produce mass gap BHs.

These findings are qualitatively aligned with observations of the top-heavy IMFs inferred in metal-poor galaxies and globular clusters \citep{Marks:2012ia,Geha2013,Gennaro18,Weatherford:2021zdf} and the work in \citet{Chruslinka2020} who highlight the importance of a galaxy/metallicity varying IMF. Alternatively, these black holes may predominantly form through dynamical capture mechanisms \citep{Antonini2022vib,Banerjee:2021wqh,Torniamenti:2024uxl} or result from the possible redshift evolution of the PISN mass scale and maximum black hole mass \citep{Mukherjee:2021rtw}. The impacts of these disparate scenarios on the mass ratio distribution and the redshift distribution may offer clues to distinguish them. 

Addressing the high-mass end of the merger-rate spectrum remains a challenge and is an active area of research. As mentioned above, the peak at $\sim 35 \, \rm M_\odot$ could be populated by a subpopulation of black holes forming in globular clusters \citep{Antonini2022vib,Banerjee:2021wqh,Torniamenti:2024uxl}. In our analysis, this bump arises due to the pair instability mechanism and hence becomes more prominent when considering a top-heavy IMF (running-slope model). However, we have treated each galaxy as a monolithic entity, neglecting diverse components such as globular clusters, star clusters, bulges, and disks. These components exhibit varying densities, metallicities and stellar populations, leading to differences in merger rates and dynamical interactions. For instance, \citet{Chowdhury:2024rjo} show that differences in metallicity between the primary and secondary components within stellar clusters lead to the formation of more massive compact objects and an increased frequency of eccentric, inspiraling binary systems. This kind of diversity could contribute to filling the mass gap through hierarchical mergers and to the formation of a more prominent peak at $\sim 35 \, \rm M_\odot$. A more granular examination of these factors, set within the framework of galactic relations, is left for future work. 

Beyond $50 \, \rm M_\odot$, our model predicts a sharp decline in the number of BBH systems, revealing the well-known mass gap caused by the pair-instability mechanism and mass loss in binary systems. Incorporating mergers in dynamically-formed BBHs originating from the single black hole population could help fill this gap. We estimate that an efficiency for these dynamical processes of $\sim 0.01\%$  could impact the high-mass end, making it more comparable to observations up to about $60~M_\odot$ in primary black hole mass. This coupled with the uncertainties in the location of the pair-instability mass gap can shift our results further more. 

We also direct the reader to previous studies on possible explanations for the high-mass end of the distribution, including population III stars \citep{Farrell:2020zju,Kinugawa:2020xws,Liu:2020lmi,Tanikawa:2020abs,Safarzadeh:2021ttc}, hierarchical mergers \citep{Sedda:2023big,Gerosa2021,Mapelli2020b,Farr2017,Antonini2022vib,Torniamenti:2024uxl}, primordial black holes \citep{DeLuca:2020sae}, and beyond the Standard Model (BSM) physics \citep{Croon:2020oga,Sakstein:2020axg,Ziegler:2020klg,Ziegler:2022apq,Croon:2023trk}.

The redshift evolution of the primary mass spectrum is investigated and we find that the positions of the $\sim 6 \, \rm M_\odot$ and $\sim 35 \, \rm M_\odot$ peaks in our model remain fixed over cosmic time. The primary change is a gradual flattening of the slope between the peaks toward higher redshifts, reflecting the larger contribution from metal-poor environments at earlier times. Such conditions favor the formation of more massive black holes relative to the low-mass high-metallicity systems that dominate at lower redshifts. This stability in peak location, coupled with metallicity-driven slope evolution, is consistent with LVK observations \citep{Fishbach:2021yvy,GWTC3pop,lallemam25}, suggesting that the underlying mass scales set by stellar physics remain constant in the isolated channel.

We also explore how our model compares to the secondary mass spectrum and mass ratio distribution. The secondary spectrum has been largely unexplored and a BBH-specific inferred rate from LVK has not been presented. Recent studies by \citet{Farah:2023swu,Sadiq:2023zee} explore inferences from LVK observations concerning the properties of secondary components. We find that our calculated secondary mass spectrum based on the astrophysical population has weaker peak structure when compared to the primary mass spectrum. For example, the peak that is prominent at $\sim 35 \, \rm M_\odot$ in $d\mathcal{R}/dM_1$ has been suppressed in $d\mathcal{R}/dM_2$. We find that both spectra receive comparable contributions from low-metallicity systems $10^{-3} \leq Z<10^{-4}$ associated with mass ratios near unity. However, $d\mathcal{R}/dM_1$ receives enhanced contributions from intermediate metallicity systems $10^{-2} \leq Z<10^{-3}$ with mass ratios distributed more broadly, leading to the suppression in $d\mathcal{R}/dM_2$. 

For the mass ratio distribution, we compare to LVK's Broken Power Law + Two Peaks (BPL+2P) and B-Spline models \citep{GWTC4pop}. We find that our calculation yields a much richer mass ratio distribution than the previous GWTC-3 results when this paper was submitted, and find structure similar to the new GWTC-4 \textit{B-Spline} model and previous non-parametric studies \citep{Edelman:2022ydv,Rinaldi:2025emt}. 
For instance, peaking at q below unity, $q\sim 0.8-0.9$ and slight kinks throughout the spectrum. Further O4 results will shed light on the high mass ratio regime.
 
In summary, our study demonstrates that a population model grounded in galactic relations, such as the galaxy stellar mass function, metallicity distribution, and star formation, can get the correct number of mergers with peaks in the mass spectrum near $\sim 10 \, \rm M_\odot$ and $\sim 35 \, \rm M_\odot$ and redshift evolution of BBHs inferred by the LVK collaboration. Our modeling suggests that for the isolated channel alone, a top-heavy IMF in low-mass metal-poor galaxies would be required to match the primary mass spectrum up to about $50 \, \rm M_\odot$. Alternately, we would need to include dynamical capture mechanisms such as those that could happen in globular clusters. Finally, our mass ratio distribution does not follow a strict power-law assumed by previous LVK O3 results, but is in agreement with the newly released LVK O4a \textit{B-Spline} model. Although the uncertainties in stellar evolution and star formation history are still large, our findings demonstrate that the bulk of the BBH mergers with primary mass below $\sim 30 \, \rm M_\odot$ observed by LVK can be explained by a simple model that averages over the observed properties of galaxies. This allows for a more robust assessment of the mismatch between the predictions and observations of black holes at the high-mass end.  

\section{Acknowledgements}
We sincerely appreciate the valuable feedback and suggestions provided by the reviewers. TBS is grateful for conversations with Floor Broekgaarden. TBS acknowledges support from the National Science Foundation Graduate Research Fellowship Program under Grant No. 1839285.  MK acknowledges support from the National Science Foundation under Grant No. 2210283.

\bibliography{mybib}{}
\bibliographystyle{aasjournal}

\begin{figure}[]
    \centering
    \includegraphics[width=0.5\textwidth]{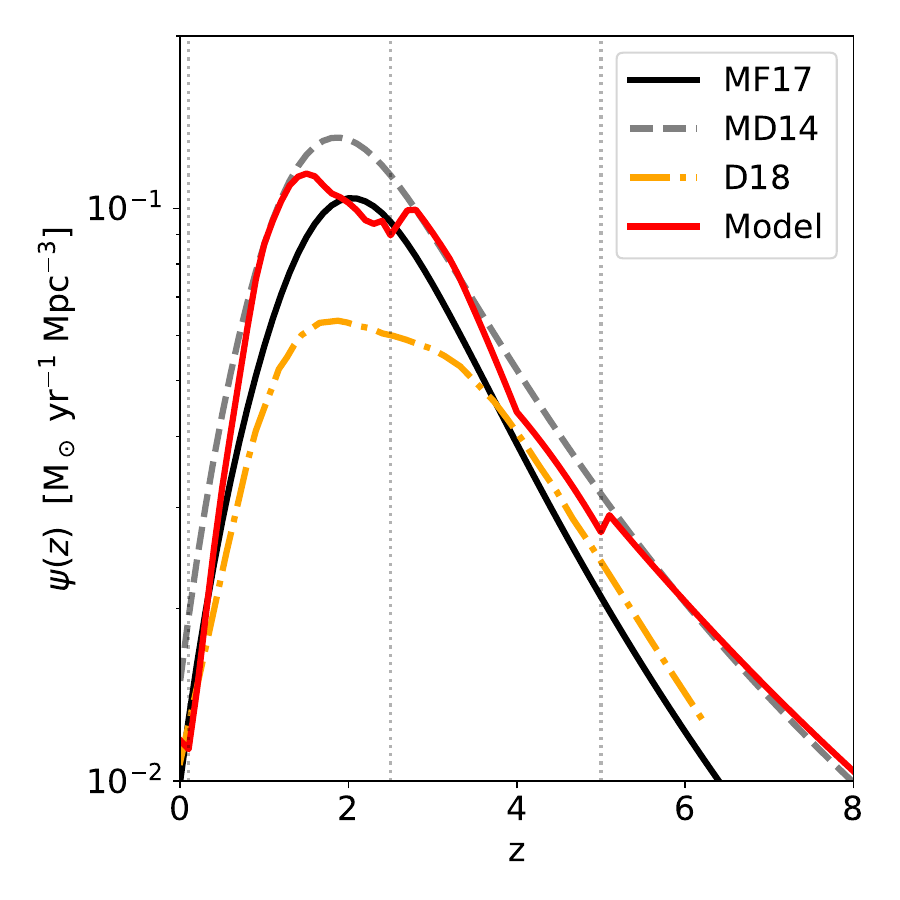}
    \caption{Cosmic star formation rate density (SFRD) derived from the convolution of the \citet{Speagle:2014loa} star-forming main sequence and the \citet{COSMOS2020} galaxy stellar mass function. The red curve shows our model prediction, compared with empirical fits from \citet{Madau:2014bja} (MD14; gray dashed), \citet{Madau2017} (M17; black solid), and \citet{Driver:2017qne} (D18;orange). Vertical gray lines mark the transitions between GSMF regimes: (1) low-redshift \citet{Baldry2012}, (2) the switch from double- to single-Schechter fits in \citet{COSMOS2020}, and (3) the onset of our high-redshift power-law extrapolation. Our model reproduces the broad SFRD peak around $z \approx 1.5$, consistent with recent analyses by D18 and \citet{behroozi_mnras2019}.
}
    \label{fig:SFRD}
\end{figure}

\appendix
\section{Star Formation Rate Density}
\label{app:mssfr}

In this appendix, we discuss how the star-forming main sequence from \citet{Speagle:2014loa}, when integrated with the star-forming galaxy stellar mass function (GSMF) from \citet{COSMOS2020}, produces the evolution of the cosmic star formation rate density (SFRD). This provides a consistency check for our adopted parameterizations of both the main sequence and the GSMF across cosmic time.

Figure~\ref{fig:SFRD} shows the resulting SFRD (red curve) compared with the empirical fits from \citet{Madau:2014bja} (MD14; gray dashed),  \citet{Madau2017} (MF17; black solid), and \citet{Driver:2017qne} (D18; orange dash-dot). The vertical gray lines indicate transitions between regimes in our adopted GSMF:
(1) the low-redshift double-Schechter fit from \citet{Baldry2012} applied at $z < 0.2$,
(2) the switch from the double- to single-Schechter fit from \citet{COSMOS2020}$,$ and
(3) the transition to the high-redshift extrapolation based on our power-law fits to the evolution of the Schechter parameters. 

Our star-formation rate density peaks at $z \sim 1.5$, shifted from MD14 and M17 but in close agreement with D18, based on the GAMA/G10--COSMOS/3D--HST compilation, and with the modeled SFRD of \citet{behroozi_mnras2019}. This indicates a slightly later turnover than the canonical $z \sim 2$. These analyses suggest that the global SFRD maximum is broader and shifted to lower redshift than previously assumed, consistent with our results and supporting our adopted framework.

\end{document}